\documentclass[10pt]{article}
\usepackage{amsmath}
\usepackage{graphicx}
\usepackage{amsfonts}
\usepackage{amssymb}
\usepackage{epsf}
\usepackage{latexsym}

\def\foo{\footnote}
\def\hat{\widehat}
\def\tilde{\widetilde}

\def\beq{\begin{equation}}
\def\eeq{\end{equation}}
\def\bea{\begin{eqnarray}}
\def\eea{\end{eqnarray}}
\def\pa{\partial}
\def\d{\textrm{d}}
\def\R{\underline{\mbox{R}}}
\def\P{\underline{\mbox{P}}}
\def\r{\underline{\mbox{r}}}
\def\ttH{\mbox{\tt H}}
\def\ttL{\mbox{\tt L}}
\def\ttD{\mbox{\tt D}}
\def\ttP{\mbox{\tt P}}

\def\C{{\mbox{C}}}

\def\cr{\mbox{\scriptsize{\bf $\mbox{ } \times \mbox{ }$}}}

\def\md{\mbox{d}}

\def\n{\mbox{N}}

\def\mA{\mbox{A}}
\def\mB{\mbox{B}}

\def\tip{\tilde{p}}
\def\tiq{\tilde{q}}
\def\tir{\tilde{r}}
\def\tis{\tilde{s}}
\def\sa{\mbox{\scriptsize a}}
\def\sb{\mbox{\scriptsize b}}
\def\sc{\mbox{\scriptsize c}}
\def\sd{\mbox{\scriptsize d}}
\def\se{\mbox{\scriptsize e}}

\def\sg{\mbox{\scriptsize g}}

\def\sk{\mbox{\scriptsize k}}
  
\def\sm{\mbox{\scriptsize m}}
\def\sn{\mbox{\scriptsize n}} 
 
\def\sp{\mbox{\scriptsize p}}

\def\sr{\mbox{\scriptsize r}}

\def\sw{\mbox{\scriptsize w}}

\def\sA{\mbox{\scriptsize A}} 
\def\sB{\mbox{\scriptsize B}}
\def\sC{\mbox{\scriptsize C}}
\def\sD{\mbox{\scriptsize D}}

\def\sF{\mbox{\scriptsize F}}

\def\sM{\mbox{\scriptsize M}} 
\def\sn{\mbox{\scriptsize N}}
\def\sN{{\cal N}} 
\def\sO{\mbox{\scriptsize O}}

\def\sT{\mbox{\scriptsize T}}

\def\eph(B){\mbox{\scriptsize emergent(LMB)}}

\def\tC{\mbox{\tiny C}}

\def\bM{\mbox{{\bf M}}}
\def\bN{\mbox{{\bf N}}}
\def\bR{\mbox{{\bf R}}}
\def\bh{\mbox{{\bf h}}}
\def\bm{\mbox{{\bf m}}}
\def\bn{\mbox{{\bf n}}}
\def\bq{\mbox{{\bf q}}}
\def\br{\mbox{{\bf r}}}
\def\bw{\mbox{{\bf w}}}
\def\bx{\mbox{{\bf x}}}
\def\bz{\mbox{{\bf z}}}
\def\sbm{\mbox{{\bf \scriptsize m}}}
\def\sbn{\mbox{{\bf \scriptsize n}}}

\def\sbA{\mbox{{\bf \scriptsize A}}}

\def\sbM{\mbox{{\bf \scriptsize M}}}
\def\sbN{\mbox{{\bf \scriptsize N}}}

\def\fp{\mbox{\sffamily p}}

\def\fr{\mbox{\sffamily r}}
\def\fs{\mbox{\sffamily s}}

\def\fA{\mbox{\sffamily A}}

\def\fE{\mbox{\sffamily E}}

\def\fH{\mbox{\sffamily H}}
\def\fI{\mbox{\sffamily I}}

\def\fL{\mbox{\sffamily L}}

\def\fP{\mbox{\sffamily P}}
\def\fQ{\mbox{\sffamily Q}}
\def\fR{\mbox{\sffamily R}}
\def\fS{\mbox{\sffamily S}}
\def\fT{\mbox{\sffamily T}}
\def\fU{\mbox{\sffamily U}}
\def\fV{\mbox{\sffamily V}}

\def\scR{\mbox{\scriptsize ${\cal R}$}}

\def\sfH{\mbox{\sffamily{\scriptsize H}}}

\def\sfL{\mbox{\sffamily{\scriptsize L}}}

\def\sfP{\mbox{\sffamily{\scriptsize P}}}

\def\sfS{\mbox{\sffamily{\scriptsize S}}}

\def\b{\underline{\mbox{B}}}
\def\p{\underline{\mbox{p}}}
\def\q{\underline{\mbox{q}}}
\def\a{\underline{\mbox{A}}}

\newcommand{\N}{{\cal N}}

\def\bn{\mbox{\bf n}}
\def\bp{\mbox{\bf p}}
\def\bA{\mbox{\bf A}}

\def\bP{\mbox{\bf P}}
\def\bM{\mbox{\bf M}}

\begin{document}
\begin{titlepage}
\vspace{.7in}
\begin{center}
\Large{\bf FOUNDATIONS OF RELATIONAL PARTICLE DYNAMICS}\normalsize

\vspace{.4in}

\large{\bf Edward Anderson$^*$} 

\vspace{.2in}

\large{\em Peterhouse, Cambridge CB2 1RD}\normalsize

\vspace{.2in}

\large{and}

\vspace{.2in}

\large{\em DAMTP, Centre for Mathematical Sciences, Wilberforce Road, Cambridge CB3 OWA.}

\end{center}

\begin{abstract}
Relational particle dynamics include the dynamics of pure shape and cases in which absolute scale 
or absolute rotation are additionally meaningful.  
These are interesting as regards the absolute versus relative motion debate as well as discussion 
of conceptual issues connected with the problem of time in quantum gravity.  
In spatial dimension 1 and 2 the relative configuration spaces of shapes are n-spheres and complex 
projective spaces, from which knowledge I construct natural mechanics on these spaces.  
I also show that these coincide with Barbour's 
indirectly-constructed relational dynamics by performing a full reduction on the latter.    
Then the identification of the configuration spaces as n-spheres and complex projective spaces, 
for which spaces much mathematics is available, significantly advances 
the understanding of Barbour's relational theory in spatial dimensions 1 and 2.  
I also provide the parallel study of a new theory for which position and scale are purely relative 
but orientation is absolute.  
The configuration space for this is an n-sphere regardless of the spatial dimension, which renders 
this theory a more tractable arena for investigation of implications of scale invariance 
than Barbour's theory itself.     

\end{abstract}

\noindent PACS: 04.60Kz. 

\mbox{ }

\vspace{5in}

\noindent$^*$ea212@cam.ac.uk

\end{titlepage}

%========================================================================================================
%========================================================================================================
\section{Introduction}
%========================================================================================================
%========================================================================================================

%========================================================================================================
\subsection{Motivation for studying relational particle models}
%========================================================================================================

General Relativity (GR) can be studied as a dynamics by splitting spacetime with respect to a family of spatial hypersurfaces 
\cite{ADM}.  
These are to have some fixed topology $\Sigma$, which I take to be a compact without boundary one for 
simplicity.    
A (rather redundant) configuration space on this is 
\noindent Riem($\Sigma) \times \mbox{Diff}(\Sigma) \times \mbox{A}(\Sigma)$ -- the values that can be 
taken by a Riemannian 3-metric $h_{\mu\nu}$ on the 3-space, by the shift $\beta_{\mu}$ (displacement in 
spatial coordinates in moving between neighbouring spatial hypersurfaces), and by the lapse $\alpha$ 
(proper time elapsed in moving between neighbouring spatial hypersurfaces).  
Then the Arnowitt--Deser--Misner (ADM) \cite{ADM} action\foo{$\bar{\sfL}$ denotes Lagrangian density.
%%%%%%%%%%%%%%%%%%%%%%%%%%%%%%%%%%%%%%%%%%%%%%%%%%%%%%%%%%%%%%%%%%%%%%%%%%%%%%%%%%%%%%%%%%%%%%%%%%%%%%%%% 
$h$ and $R$ are the determinant and Ricci scalar of $h_{\mu\nu}$.  
$\Lambda$ is the cosmological constant.  
I use $(\mbox{ },\mbox{ })_{\sbA}$ for the 
inner product with respect to the array $\bA$, with corresponding norm $||\mbox{ }||_{\sbA}$.
When $\bA$ is the identity matrix, it is dropped from the notation.  
In the present use, the array $\bM$ is the GR configuration space metric, with components 
$\sM^{\mu\nu\rho\sigma} = h^{\mu\rho}h^{\nu\sigma} - h^{\mu\nu}h^{\rho\sigma}$ and determinant $\sM$.  
This is the inverse of the undensitized DeWitt \cite{DeWitt67} supermetric $\bN$, which has components
$\mbox{\scriptsize N}_{\mu\nu\sigma\rho} = h_{\mu\nu}h_{\rho\sigma} - \frac{1}{2}h_{\mu\nu}h_{\rho\sigma}$. 
$\lambda$ is label-time and the dot denotes $\pa/\pa\lambda$.
$\pounds_{\beta}$ is the Lie derivative with respect to the vector field $\beta_{\mu}$.}   
%%%%%%%%%%%%%%%%%%%%%%%%%%%%%%%%%%%%%%%%%%%%%%%%%%%%%%%%%%%%%%%%%%%%%%%%%%%%%%%%%%%%%%%%%%%%%%%%%%%%%%%%%
\beq
\fI_{\sA\sD\sM} = \int\d\lambda\int\d^3x\bar{\fL}_{\sA\sD\sM} = 
\int\d\lambda\int\d^3x\sqrt{h}\alpha
\left\{
\frac{\fT_{\sA\sD\sM}}{\alpha^2} + R - 2\Lambda
\right\} \mbox{ } , 
\eeq
where
\beq
\fT_{\sA\sD\sM} = \frac{1}{4}||\dot{\bh} - \pounds_{\beta}\bh||^2_{\sbM} = 
\frac{1}{4}\mbox{M}^{\mu\nu\rho\sigma}
\{\dot{h}_{\mu\nu} - \pounds_{\beta}h_{\mu\nu}\}\{\dot{h}_{\rho\sigma} - \pounds_{\beta}h_{\rho\sigma}\} 
\mbox{ } 
\mbox{ } , 
\label{ADMac}
\eeq 
can be constructed by likewise splitting the usual spacetime Einstein--Hilbert action of GR.  
The lapse and shift are not dynamical variables as their conjugate momenta are zero, so it is 
evident that all the physics lies within the `metrodynamics' (dynamical evolution of $h_{\mu\nu}$).  
Variation with respect to $\beta_{\mu}$ produces the momentum constraint\foo{$D_{\mu}$ is the spatial 
covariant derivative and $\pi^{\mu\nu}$ is the momentum conjugate to $h_{\mu\nu}$.}
\beq
{\cal H}_{\mu} \equiv - 2D_{\nu}{\pi^{\nu}}_{\mu} = 0 \mbox{ } ,   
\label{Mom}
\eeq
and variation with respect to $\alpha$ produces the Hamiltonian constraint, 
\beq
{\cal H} \equiv \frac{1}{\sqrt{h}} ||\pi||^2_{\sbN} - \sqrt{h}\{R - 2\Lambda\} = 0 \mbox{ } .
\label{Ham}
\eeq
Variation with respect to $h_{\mu\nu}$ provides the evolution equations, which straightforwardly 
propagate the above constraints.
The momentum constraint (\ref{Mom}) is interpretable as the geometrically-clear restriction on the 
`metrodynamics' \cite{Battelle} that the coordinate grid information in the metric is redundant rather 
than physical.  
Thus the physics is contained within the remaining, `geometrical shape' information in the metric, 
and thus GR is, more specifically, a {\it geometrodynamics} \cite{Battelle, DeWitt67} on the quotient 
configuration space superspace($\Sigma$) = Riem($\Sigma$)/Diff($\Sigma$) \cite{Battelle}.
This is less redundant (it is still partly redundant because the Hamiltonian constraint has not yet been 
addressed).  
The dynamical objects are the slices' 
geometry (i.e. their shape as opposed to how a coordinate grid is painted on that shape).  
One way of viewing geometrodynamics (which extends to fundamental matter sources) is as a spatially 
relational theory or arbitrary 3-diffeomorphism corrected theory (see e.g. \cite{RWR, Phan}).  
These frame corrections appear solely as corrections to the velocities in the action.  
This has a counterpart for particle models with such as arbitrary translation, rotation and scale 
frame corrections: relational particle models (RPM's) \cite{BB82, B03}.

Quantum geometrodynamics has a notorious problem of time \cite{DeWitt67, K81, PW83, UW89, K91, K92, I93, 
B94I, B94II, K99, EOT, 06II} because `time' takes a different meaning in GR and in ordinary quantum 
theory.  
One notable manifestation of this is that a frozen (i.e. timeless or stationary) Schr\"{o}dinger equation 
arises therein: the quantum counterpart of \ref{Ham} is the Wheeler--DeWitt equation\foo{The 
%%%%%%%%%%%%%%%%%%%%%%%%%%%%%%%%%%%%%%%%%%%%%%%%%%%%%%%%%%%%%%%%%%%%%%%%%%%%%%%%%%%%%%%%%%%%%%%%%%%%%%%%%%
inverted commas denote that the WDE has additional technical problems: there are 
operator-ordering ambiguities (the ordering I give here is the Laplacian one, see e.g. \cite{HP86, 07II} 
for motivation) and regularization is required, while what functional differential equations mean 
mathematically is open to question.}  
%%%%%%%%%%%%%%%%%%%%%%%%%%%%%%%%%%%%%%%%%%%%%%%%%%%%%%%%%%%%%%%%%%%%%%%%%%%%%%%%%%%%%%%%%%%%%%%%%%%%%%%%%
\beq
\hat{\cal H}\Psi = 
- 
{\hbar^2}
`\frac{1}{\sqrt{M}}\frac{\delta}{\delta h^{{\mu\nu}}}
\left\{
\sqrt{M}N^{\mu\nu\rho\sigma}\frac{\delta\Psi}{\delta h^{{\rho\sigma}}}
\right\}\mbox{'}
-  
\sqrt{h}R\Psi 
+ {\sqrt{h}2\Lambda   }\Psi = 0
\label{WDE} \mbox{ } .  
\eeq
The problem of time is unresolved for GR; many conceptual strategies have been put 
forward to resolve it but none to date work when examined in detail.  

\noindent 1) There are approaches that involve finding a fundamental time for the full theory at the 
classical level.  
For example, one could seek for such a time by canonically transforming the geometrodynamical variables 
to new variables among which an explicit and genuinely time-like time variable is isolated out.  
One candidate time of this form is {\it York time}.  
This is a `dilational object' built out of the gravitational quantities alone: it is proportional to 
$h_{\mu\nu}\pi^{\mu\nu}/{\sqrt{h}}$ \cite{York72, YorkTime, K81, K92, I93}.  
A different possibility is that adjoined `reference' matter fields such as Gaussian reference fluid, 
dust or null dust, could themselves provide a time \cite{K92, I93, New}.  

\noindent 2) There are strategies in which time is capable of being emergent in the quantum regime 
despite not always being present at the fundamental level.  
One example of this is superspace time: superspace is indefinite and so a time-like notion exists for it 
too.  
Another example is the semiclassical approach \cite{DeWitt67, SemiclConcat, HallHaw}, in 
which a time emerges in the semiclassical regime.    
This is additionally a useful framework for discussing the origin 
\cite{HallHaw} of galaxies and cosmic microwave background perturbations within semiclassical scheme, 
for which one needs to study spatially-located fast light degrees of freedom that are coupled to global 
slow heavy degees of freedom such as the size of the universe.

\noindent 3) There are also timeless records strategies \cite{PW83, GMH, B94I, B94II, EOT, H99, HT, 
Hallioverlap, ARec, Dist}, in which what is primary is correlations between localized 
subsystems of a single present, from which a semblance of dynamics or history is to be constructed. 

\noindent(Finally, there are also approaches in which it is the histories that are primary 
\cite{GMH, Hartle}.)

The structure of the GR configuration space plays an important underlying role in the above problem
of time investigations.  
Superspace($\Sigma$) for $\Sigma$ compact without boundary has been studied e.g. in \cite{DeWitt67, 
Superspaceconcat}, which revealed a number of its topological space, metric space, differential 
structure and geometrical properties (then e.g. the superspace time approach follows from 
the natural metric on superspace being indefinite, while records theory is built on the study 
of measures of distance between subconfigurations).   
Because the Hamiltonian constraint (\ref{Ham}) remains unaddressed at this stage, 
the information in superspace is not purely physical.  
The restriction due to (\ref{Ham}) does not admit a straightforward geometrical interpretation.  
It restricts one to 2/3 of superspace.  
A geometrically natural 2/3 of superspace is conformal superspace($\Sigma$) = 
superspace($\Sigma$)/Conf($\Sigma$) (studied e.g. in \cite{York72, CS}) for Conf($\Sigma$) the group of 
conformal transformations associated with the maximal condition \cite{Max} 
\beq
\pi \equiv h_{\mu\nu}\pi^{\mu\nu} = 0 \mbox{ } , 
\label{Max}
\eeq 
or the volume-preserving conformal transformations associated with the constant mean curvature condition 
\beq
\pi/\sqrt{h} \equiv h_{\mu\nu}\pi^{\mu\nu}/\sqrt{h} = \mbox{const} \mbox{ } 
\label{CMC} \mbox{ } .  
\eeq
One can see that this mathematics is related to the York time approach.
The above 2/3, however, might not bear any direct relation with the 2/3 of superspace picked out by the 
Hamiltonian constraint itself.   
Also, the geometrical nature of superspace and conformal superspace is extremally complicated, 
which places limitations on what insights one can get from their study.

The notorious frozen formalism aspect of problem of time in quantum GR 
stems from the homogeneous quadraticity in the momenta of the GR Hamiltonian constraint, 
a feature which can be emulated more freely of technical complications in simpler cases.
Thus toy models are often used to conceptualize about it \cite{K91, K92, I93, EOT}.    
Similar arguments make toy models a sensible starting point for other tough conceptual such as with 
quantum mechanical closed universes \cite{PW83, Hartle, H99, HT, Hallioverlap} and how one might 
envisage and find (enough) observables \cite{perenniallit, K92} in quantum GR. 
While minisuperspace (see e.g. the review \cite{Wiltshire}) is a commonly-studied toy model, it is not 
very useful for some purposes because it treats all points in space as behaving in exactly the 
same way.  
This amounts to 1) having no nontrivial linear momentum constraints [an important source of complications with 
uplifting problem of time resolving strategies to (more) full GR].  
2) There being no meaningful notion of subsystem 
that is localized in space, which is needed for e.g. the records theory approach and some aspects of 
the semiclassical approach. 
This paper's {\it relational particle dynamics} \cite{BB82, B03} toy models are a class of toy models 
that do incorporate both of features 1) and 2) among the ways in which they resemble GR, so they are a 
useful class \cite{RPMQuant, K92, B94I, EOT, Paris, 06II, SemiclI, 07I, 07II, SemiclIII, ARec, Dist, 
AngleDep} of toy models for records and semiclassical schemes. 
(they also have a dilational internal time that parallel GR's York time.)   
They do not however serve for all POT strategies: e.g. not for the superspace time approach or the 
reference matter field time approach (which is also `orthogonal' to this paper's approach to 
geometrodynamics in involving phenomenological rather than fundamental matter).     
The line of development in this paper focuses on the underlying structure of the configuration spaces 
for RPM's, and will then proceed via a second paper \cite{07II} in which I quantize the present 
manuscript's toy models.

RPM's also have a separate life as formalisms opposite to the ``absolutist" development of mechanics. 
In particular, Barbour--Bertotti theory is directly relevant \cite{BB82, Comments, buckets, Jbook} to the 
absolute or relative motion debate \cite{AORM, buckets, Jbook}, as it is a relational (Leibnizian, Machian)  
formulation of mechanics and in agreement with a subset of Newtonian mechanics (the zero total angular momentum universes).
Its 3-particle subcase, {\it triangleland}, is already a useful example of relational motion, 
featuring e.g. in \cite{EOT}.  
Barbour's dilation-invariant theory is likewise of interest: while Barbour--Bertotti theory is a 
dynamics of shape and size, Barbour theory is a dynamics of pure shape. 
Thus it is interesting to point out that 1) in this paper I obtain the models by a line of 
thought that is different from the one that Barbour used in originally finding these models. 
He worked with redundant variables while I work on the configuration space of reduced variables, 
and yet the two a priori different types of theories that result coincide, as I demonstrate in Sec 4). 
2) I obtain an extra RPM theory that hadn't been considered before.      
Barbour's theory \cite{B03, 07I} presents a possible explanation of departures from 
standard gravitational physics at the larger relative scales as following from a simple underlying physical 
principle \cite{B03, 07I}.   
This has potential for astrophysical interest, e.g. whether it could serve as a simple theoretical 
model for galaxy rotation curves while furbishing solar system physics that is consistent with 
observation.  
This paper's new dilation-invariant but absolute orientation theory could be used to 
address this possibility free of the substantial complications incurred by quotienting out 3-d rotations. 
%(which also makes it simpler to quantize).

Finally, RPM's are also interesting examples in their own right as regards applying quantization 
techniques and investigating quantum properties \cite{RPMQuant, 06I, 07II}. 
[These studies are also prerequisites for many parts of the studies of relational dynamics as toy models 
toward understanding conceptual issues in quantum GR.]   
RPM models are subtle enough for operator ordering issues (see e.g. \cite{HP86}) and global issues 
(see e.g. \cite{I84}) to be relevant.

I begin by expanding on the abovementioned difference of line of thought. 
themselves of interest because the POT issues are not yet understood.

%========================================================================================================
\subsection{Quite a general dynamical approach to classical physics}
%========================================================================================================

Given some notion of space, one is to consider some {\it configuration space} that is compatible with it. 
A configuration space consists of the set of different values that can be taken by a set of base 
objects, e.g. particle positions (see \cite{Lanczos} for a clear exposition of this case), 
inter-particle relations, the values at each point of continuous extended objects\foo{In 
%%%%%%%%%%%%%%%%%%%%%%%%%%%%%%%%%%%%%%%%%%%%%%%%%%%%%%%%%%%%%%%%%%%%%%%%%%%%%%%%%%%%%%%%%%%%%%%%%%%%%%%%%
dimension d, continuous extended objects cover fields (values everywhere in space) and objects that have their own 
separate notion of space of extent, i.e. membranes that have values on surfaces of some dimension 
between $d - 1$ and $1$, in which last case they are termed strings.} 
%%%%%%%%%%%%%%%%%%%%%%%%%%%%%%%%%%%%%%%%%%%%%%%%%%%%%%%%%%%%%%%%%%%%%%%%%%%%%%%%%%%%%%%%%%%%%%%%%%%%%%%%%
or geometrical objects, though this may further be augmented from a set by bringing in topological space, 
metric space and geometrical structures.    
The set of base objects in question is allowed to be redundant (e.g. coordinate redundancy, gauge 
redundancy), i.e. some base objects can be partly or totally empty of physical content.  
This kind of description is used because it is often impractical or beyond current technical 
understanding to work with just physical base objects.

From one's base objects, one then constructs natural compound objects (these may include velocities, 
spatial derivatives and contracted objects), perhaps subject to some limitations (from implementating 
physical or philosophical principles, or purely mathematical simplicity postulates), and assemble a 
scalar Lagrangian or Lagrangian density from these.  
This is then integrated over whatever notion of time and space of extent are appropriate to form an 
action.
Given an action, one can define the momenta conjugate to a set of configuration space coordinates and 
see if there are any inter-relations among these due to the form of the action (primary constraints).  
Variation of the action with respect to the base objects provides perhaps some secondary constraints, 
and some evolution equations.
Even more constraints may arise by the requirement that the evolution equations propagate the 
constraints (the Dirac method \cite{Dirac}).  

\mbox{ }

\noindent{\bf Postulate 1 (configurational relationalism).} One may consider that there is a 
group $G$ of motions that are physically redundant.  
This group may contain spatial and (or) internal motions, so it is a generalization of spatial 
relationalism as well as a way of thinking about gauge theory.    

\mbox{ } 

\noindent{\bf Indirect implementation of Postulate 1}.  
The velocities of the base quantities pick up arbitrary $G$-frame corrections.  
(This is usually explained in terms of symmetry requirements on the Lagrangian or in terms of the 
appending part of Dirac's procedure \cite{Dirac} in the Hamiltonian formalism followed by Legendre 
transformation to the Lagrangian formalism.)    
Thus gauge auxiliaries feature in the action.  
Then variation with respect to these then produces constraints which implement the physical irrelevance 
of $G$: each such constraint takes out both one degree of freedom in $G$ and one degree of freedom in 
$\fQ$, so that the physical content is embodied by the quotient configuration space $\fQ/G$.  

\mbox{ } 

\noindent{\bf Parametrization Procedure} One may adjoin the original notion of time's time variable to 
the configuration space 

\noindent
$\fQ \longrightarrow \fQ \times T$ by rewriting one's action in terms of a label-time parameter 
(see e.g. \cite{Lanczos}).  

\mbox{ } 

%=======================================================================================================
\noindent{\bf Example 1: the ADM formulation of GR}  
%=======================================================================================================

\mbox{ } 

\noindent The above postulates can be taken to underly the formulation of GR in Sec 1.1.
The notion of space here is in the background topology $\Sigma$ and the incipient notion of 
configuration space is Riem($\Sigma$).  
The adjunction of Diff($\Sigma$) is an example of the indirect implementation of configurational 
relationalism, corresponding to the coordinatization of $\Sigma$ being held to be physically irrelevant.   
The adjunction of $\mbox{A}(\Sigma)$ is a rather special example of the parametrization procedure.  
From these metric, shift and lapse base objects, one can construct the compound objects of metric 
geometry and the 3-diffeomorphism-corrected (i.e. shift-corrected) metric velocities.  
One can then write down many different actions on this configuration space.
Among these, the ADM action (\ref{ADMac}) can be taken to follow either from ADM's decomposition of the 
spacetime action or from a sufficiently long list of principles and simplicity assumptions (though in 
fact the form of the spacetime action itself depends on a number of simplicity assumptions, so some 
entirely mathematical assumptions, such as about the highest-order derivatives that are to feature 
in the action, have to be made at some stage in arriving at the above action).

The context in which the RPM's arose is a somewhat different formulation of physics to this Subsection's,  
which I next present.   
[It should be noted that these differences do not spoil the Hamiltonian form of the subsequent 
constraint equations.]

%========================================================================================================
\subsection{Barbour-type dynamical approach to classical physics}
%========================================================================================================

Replace the indirect implementation of Postulate 1 by the more general alternative 

\mbox{ }

\noindent{\bf Barbour-type implementation of Postulate 1}: consider arbitrary $G$-frame forms for the 
base objects and all the compound objects.  
Now, in commonly-encountered examples (which include all the examples in this Paper), 
$\fQ$ and $G$ happen to be sufficiently compatible that corrections to the objects themselves do 
not show up\foo{See \cite{ABFO, ABFKO, FEPII} for discussion of more 
complicated cases for which this is not the case.}, but setting up the arbitrary $G$-frame and taking 
the time derivative do not commute, so that the velocity of the original auxiliary variable does appear 
in a correction to the velocity of each base object in $\fQ$.  
[This is a derivation \cite{ABFO, Lan} of Barbour's `best matching' in the first sense in which he uses 
this expression (see e.g. \cite{BB82, RWR}).]  

\mbox{ } 

\noindent Barbour also does not trust the parametrization procedure because the time variable is 
extraneous to the configuration space.  
He would start rather with 

\mbox{ } 

\noindent{\bf Postulate 2 (temporal relationalism).} Time is but a label in sufficiently 
general and fundamental physics.  

\mbox{ } 

\noindent This is to be implemented as follows: 
`actions are to be built to be manifestly reparametrization invariant'. 
Using as base objects the {\it instant } I such that $\dot{\mbox{I}} = \alpha$ and {\it frame} 
$\mbox{F}_{\mu}$ such that $\dot{\mbox{F}}_{\mu} = \beta_{\mu}$ \cite{FEPI}, I note that, nevertheless, 
the parametrization procedure 2 is a subimplementation of this.   
Note that while the lapse and shift were multipliers, these new instant and frame variables are 
cyclic coordinates.  
That this does not affect the outcome of the variational procedure (which hinges on these variables 
nevertheless being auxiliary and hence freely prescribable at the endpoints of variation) is the subject 
of \cite{FEPI}.    
However, using either lapse or instant is still in tension as regards Barbour's issue with 
extraneousness, while there is another subimplementation that is not,   

\mbox{ } 

\noindent{\bf Barbour-type subimplementation of Postulate 2}: actions are to be built to be manifestly reparametrization invariant 
without extraneous time variables.  

\mbox{ } 

\noindent[Two further goals of the Barbour approach are as follows.  
One is also to search for a minimizer to establish the least incongruence between adjacent physical 
configurations (that is the second sense in which Barbour uses the expression `best matching').   
One is also interested in obtaining geodesics on the reduced configuration space that uniquely represent 
the dynamical motion.]
 
\mbox{ }

%========================================================================================================
\noindent{\bf Example 2: BFO-A formulation of GR}   
%========================================================================================================

\mbox{ }

\noindent Again, one has a topological notion of 3-space with some fixed topology $\Sigma$, which
I take to be a compact without boundary topology.  
Then a (rather redundant) configuration space on this is Riem($\Sigma) \times \mbox{ Diff}(\Sigma)$, 
corresponding to the values that can be taken by a Riemannian 3-metric $h_{\mu\nu}$ on the 3-space, and 
by the frame $\mbox{F}_{\mu}$ (the spatial coordinates themselves); again, the adjunction of 
Diff($\Sigma$) corresponds to the coordinatization of $\Sigma$ being held to be physically irrelevant.   
From these base objects, one can construct the frame-corrected objects of the spatial metric geometry.  
As these transform well under the 3-diffeomorphisms of the spatial 3-metric geometry, the Diff$(\Sigma)$ 
corrections are manifest only as corrections to the metric velocities. 
One then assembles the action \cite{RWR, Lan, ABFO, Phan}
\beq
\fI_{\sB\sF\sO{-}\sA} = \int\d\lambda\int\d^3x\bar{\fL}_{\sB\sF\sO{-}\sA} = 
\int\d\lambda\int\d^3x\sqrt{h}\sqrt{\fT_{\sB\sF\sO{-}\sA}\{R - 2\Lambda\}}
\mbox{ } , \mbox{ } \fT_{\sB\sF\sO{-}\sA} = ||\dot{h} - \pounds_{\dot{\sF}}h||^2_{\sM} \mbox{ } ,
\eeq
based on manifest reparametrization invariance without extraneous time variables, on the usual kind of 
simplicity postulates and the observation that the Dirac procedure prevents other likewise simple 
choices for kinetic term $\fT$ from working \cite{RWR, Lan, Phan, Lan2}.  
One can also obtain this by doing the instant-frame version of the ADM split \cite{FEPI} on the 
Einstein--Hilbert action and then eliminating the velocity of the instant by Routhian reduction 
(a move which directly parallels \cite{FEPI} Baierlein, Sharp and Wheeler's \cite{BSW} elimination 
of the lapse Lagrange multiplier).

In this approach, the Hamiltonian constraint now arises as a primary constraint, while the momentum 
constraint arises similarly to before as a secondary constraint from variation with respect to an 
auxiliary (now the frame rather than the shift).  
Again, propagation by the evolution equations gives no further constraints.

There is also a parallel to this approach in which volume-preserving conformal transformations 
are also considered to be a priori physically meaningless, which encodes constant mean curvature sliced 
GR \cite{CMC, CMCapps} from a variational principle \cite{ABFKO}.

%========================================================================================================
\subsection{Relational particle dynamics}
%========================================================================================================

Following the Barbour version of the above scheme gives rise to relational particle dynamics models.  
The incipient notion of space is {\it Absolute space} $\fA(\d) = \mathbb{R}^{\sd}$.  
In each case the incipient {\it configuration space} 

\noindent $\fQ(\N, \d) = \langle\N$ labelled possibly 
superposed material points in $\mathbb{R}^{\sd}\rangle$.  
One then considers a group $G$ comprising any combination of: absolute translations, absolute rotations 
and absolute scales to be physically meaningless by adjoining $G$ to $\fQ(\N,\d)$.  
In each case one considers a {\it Jacobi-type action} \cite{Lanczos} (these are always manifestly reparametrization 
invariant and free of time variables extraneous to the configuration space):  
\beq
\fI = \int\d\lambda \fL = 2\int\d\lambda\sqrt{\fT\{\fU + \fE\}} \mbox{ } ,   
\label{action}
\eeq
where the kinetic term $\fT$ is homogeneous quadratic in the velocities.  
$\fU$ is minus the potential energy $\fV$ and $\fE$ is the total energy.
Note that each such action is indeed equivalent to the more well known Euler--Lagrange actions
\beq
\fI = \int\d\lambda \fL = \int\d\lambda\{\fT - \fV\} \mbox{ } , 
\label{Lagaction}
\eeq
(see \cite{Lanczos} for (\ref{action}) $\Rightarrow$ (\ref{Lagaction}) by Routhian reduction and 
\cite{SemiclI} for (\ref{Lagaction}) $\Rightarrow$ (\ref{action}) by the emergence of a lapse-like or 
instant-like quantity).

$\fV$ and $\fT$ are constructed from the $G$-frame corrected basic objects.  
As translations, rotations and scalings are compatible with the vectorial notion of particle positions,  
arbitrary $G$-frame corrections show up only as corrections to the velocities.

\noindent Actions of this form lead to each such theory having as a primary constraint a quadratic 
energy constraint that is analogous with the Hamiltonian constraint of GR in giving rise to a frozen 
formalism problem.   
Variation with respect to the adjoined $G$-auxiliary variables produces constraints which ensure 
that one passes from $\fQ(\N, \d) \times G$ to the quotient space $\fQ(\N, \d)/G$.  
In each case considered, the evolution equations from variation with respect to the particle position 
coordinates that make up the $\fQ(\N, \d)$ propagate the above constraints without producing any more 
constraints.  

\mbox{ } 

\noindent{\bf Example 3 (Barbour--Bertotti theory)}. 
\cite{BB82} is the original reference; see \cite{B94I, EOT, Comments, RPMQuant, LBGGM, 06I, 06II, 
SemiclI, AngleDep} for developments). 
Here $G$ is Eucl(d), the Euclidean group of translations and rotations, so that overall translations and 
rotations of the model universe are taken to be meaningless.  
I denote the associated auxiliary variables by $\a$ and $\b$.  
An action for this theory is then (\ref{action}) with\foo{I use calligraphic indices for particle 
position labels from 1 to $\N$ and capital indices for relative position labels from 1 to N = $\N$ -- 1.
I sometimes use underling in place of spatial indices, and bold in place of both spatial indices and position labels. 
$\bm$ is the array with components $m_{\cal A}\delta_{\cal AB}$ for $m_{\cal A}$ the ${\cal A}$th 
particle's mass.
$\bn$ is the inverse of this array. 
$\mbox{\boldmath${\mu}$}$ is the array with components $\mu_I\delta_{IJ}$ for $\mu_I$ the Ith Jacobi mass 
\cite{Marchal}.  
$\mbox{\boldmath${\nu}$}$ is the inverse of this array.} 
\beq
\fT = \frac{1}{2}||\dot{\bq} - \dot{\a} - \dot{\b} \cr \bq||_{\sbm}^2  \mbox{ } , 
\label{18}
\eeq
\beq
\fV = \fV(\|\q_{\cal A} - \q_{\cal B}\| \mbox{ alone}) \mbox{ }  ,
\label{19}
\eeq

\noindent where the $\underline{\mbox{q}}_{\cal A}$ are particle positions. 
Then the conjugate momenta are  
\beq
\p^{\cal A} =  \delta^{\cal AB}m_{\cal B}\{\dot{\q}_{\cal B} - 
\dot{\a}_{\cal B} - \dot{\b} \cr \q_{\cal B}\}/\dot{\mbox{I}} \mbox{ } 
\eeq
for $\dot{\mbox{I}} = {\fT}/\{\fU + \fE\}$.  
These are seen to obey a primary constraint, the quadratic energy constraint 
\beq
\ttH \equiv \frac{1}{2}||\bp||_{\sbn}^2 + \fV - \fE = 0 \mbox{ } .
\eeq
Variation with respect to $\a$ and $\b$ yields respectively the secondary constraints
\beq
\underline{\ttP} \equiv \sum_{\cal A}\p^{\cal A} = 0 \mbox{ } , \mbox{ } 
\label{ZM}
\eeq
\beq
\underline{\ttL} \equiv \sum_{\cal A}\q_{\cal A} \cr \p^{\cal A} = 0 \mbox{ } ,  
\label{ZAM}
\eeq
i.e. linear zero total momentum and zero total angular momentum constraints.  
These various constraints may be recast in terms of Jacobi coordinates $\R_A$ which are \cite{Marchal} 
`diagonalizing' linear combinations of the independent relative particle positions 
$r_{\cal AB} = q_{\cal A} - q_{\cal B}$.
$\ttP = 0$ is trivially eliminated by this change of coordinates, while $\ttH$ and $\underline{\ttL}$ 
become the similar-looking expressions  
\beq
\ttH \equiv \frac{1}{2}||\bR||_{\mbox{\scriptsize\boldmath${\mu}$}}^2 + \fV - \fE = 0 \mbox{ } , 
\label{Jacen}
\eeq 
\beq
\underline{\ttL} = \sum_{A = 1}^{\sn}\R_A \cr \P^A = 0 \mbox{ } ,
\label{Jacam}
\eeq
where the Jacobi momenta are $\P^A =  \delta^{AB}\mu_{B}\{\dot{\R}_{B} - 
\dot{\b} \cr \R_{B}\}/\dot{\mbox{I}}$.
It is (\ref{Jacam}) which plays the role of a nontrivial analogue of the GR momentum constraint.   

\mbox{ } 

\noindent{\bf Example 4 (Barbour's theory)} \cite{B03} is the original reference; for developments of 
this see \cite{Piombino, 06II, 07I} and below.  
Here $G$ is Sim(d), the Similarity group of translations, rotations and dilatations so that overall 
translations, rotations and dilatations of the model universe are taken to be meaningless.  
I denote the associated auxiliary variables by $\a$, $\b$ and $\C$.   
An action for this theory is then (\ref{action}) with 
\beq
\fT = \frac{1}{2j}\left|\left|\dot{\bq} - \dot{\a} - \dot{\b} \cr \bq + \dot{\C}\bq|\right|\right|_{\sbm}^2  
\mbox{ } , 
\label{18b}
\eeq
\beq
\fV = \fV(\|\q_{\cal A} - \q_{\cal B}\| \mbox{ alone}) \mbox{ }   
\label{19b}
\eeq 
which is additionally homogeneous of degree zero.  
$j = ||\bq||^2_{\sbm}$, the total moment of inertia of the system.

The corresponding momenta are 
\beq
\p^{\cal A} = \delta^{\cal AB}m_{\cal A}\{\dot{\q}_{\cal B} - \dot{\a} + \dot{\C}\q_{\cal B}\}/{j\dot{\mbox{I}}} 
\mbox{ } .  
\label{mom4}
\eeq
These again obey an energy constraint, now of form
\beq
\frac{j}{2}||\bp||^2_{\sbn} + \fV = \fE \mbox{ } ,  
\label{Dilen}
\eeq
momentum and angular momentum constraints of the same forms as above (\ref{ZM}), (\ref{ZAM}),  
and a new zero dilational momentum constraint which arises from variation with respect to $\C$, 
\beq
\ttD \equiv (\bq \cdot \bp) = 0 \mbox{ } .  
\label{ZDM}
\eeq
Passing to Jacobi cordinates, the zero momentum constraint is again absorbed, while the other 
constraints take the forms  
\beq
\ttH \equiv \frac{J}{2}||\bR||_{\mbox{\scriptsize\boldmath${\mu}$}}^2 + \fV - \fE = 0 \mbox{ } , 
\label{JacenI}
\eeq 
(\ref{Jacam}) and
\beq
\ttD = (\bR \cdot \bP) = 0 \mbox{ } , 
\label{Dil}
\eeq
where now the Jacobi momenta are $\mbox{\underline{P}}^A = \delta^{AB}\mu_B\{\dot{\R}_{B} - 
\dot{\b} \cr \R_{B} + \dot{\C}\R_{B}\}/J\dot{\mbox{I}}$ and $J = ||\bR||^2_{\mbox{\scriptsize\boldmath${\mu}$}}$, the 
moment of inertia of the system expressed in Jacobi coordinates.  
This last constraint is manifestly analogous to the GR maximal slice condition (\ref{Max}) [while 
$(\bR \cdot \bP)$ serves as a notion of internal time \cite{06II} in Barbour--Bertotti theory, much as the mean curvature 
(\ref{CMC}) serves as the York time notion in GR].  

\mbox{ } 

\noindent{\bf Example 5 (A new theory)}.  
Here $G$ is the group of translations and dilatations, so that overall translations and dilatations of 
the model universe are taken to be meaningless, but absolute rotations retain physical significance. 
I denote the associated auxiliary variables by $\a$ and $\C$.  
This is augmented by the group of translations and dilatations so that overall translations and 
dilatations of the model universe be meaningless.
An action for this theory is (\ref{action}) with 
\beq
\fT = \frac{1}{2}\left|\left|\dot{\bq} - \dot{\a} + \dot{\C}\bq\right|\right|_{\sbm}^2  \mbox{ } ,
\eeq
and $\fV = \fV(||q_{\cal A} - q_{\cal B}||)$ alone which is additionally homogeneous of degree zero.  
This is a new theory for d $>$ 1 (while it coincides with B theory for d = 1, as rotations are then 
trivial).

The corresponding momenta are
\beq
\p^{\cal A} =  
\delta^{\cal AB}m_{\cal A}\{\dot{\q}_{\cal B} - \dot{\a} + \dot{\C}\q_{\cal B}\}/j\dot{\mbox{I}} 
\mbox{ } .  
\eeq
These obey energy, momentum and dilational momentum constraints as above: (\ref{Dilen}), (\ref{ZM}) 
and (\ref{ZDM}).  
On passing to Jacobi coordinates, this leaves one with (\ref{JacenI}) and (\ref{Dil}), now for Jacobi 
momenta of form $\mbox{\underline{P}}^A = \delta^{AB}\mu_B\{\dot{\R}_{B} - \dot{\C}\R_{B}\}/
J\dot{\mbox{I}}$.

%========================================================================================================
\subsection{Outline of the rest of this Paper}
%========================================================================================================

It is known that at least some cases of RPM theories can furthermore be formulated otherwise by 
reduction \cite{LBGGM, 06I, 06II, 07I}, i.e. by elimination of some of the auxiliary velocities  
$\dot{\b}$ and $\dot{\C}$.  
However, in this paper, rather, I start (Sec 2) from the configuration space and build upwards using 
what natural topological space, metric space and Riemannian geometry that one has on these configuration 
spaces, based on Kendall's work \cite{KendallandKendall, Kendall84, Kendallbook}, and thus arrive at 
(Sec 3) likewise natural mechanics directly implemented on these spaces.  
That involves replacing the above indirect implementations of Postulate 1 by the  

\mbox{ } 

\noindent{\bf direct implementation of Postulate 1}: work on the quotient spaces $\fQ/G$ themselves.  

\mbox{ } 

\noindent I succeed thus in building arbitrary particle number mechanics in dimensions 1 and 2, and in 
arbitrary dimension for the case with absolute rotation but no absolute scale. 
Appendix A provides the geometrical computations needed to make my equations for these mechanics 
explicit.     
I then show (Sec 4) that these are the same theories that one obtains from fully reducing the 
Barbour-type theories.  
Thus this Paper points out that there is available (and presents) an essentially complete configuration 
space study available for the above-listed cases of these relational particle mechanics theories.  
In this regard, Appendix A provides further useful geometrical properties and Appendix B provides useful 
topological properties.  
These are very useful as regards quantizing these theories, and thereby as regards using them as 
toy models for semiclassical and records theory approaches (and yet other approaches) to the problem of 
time in quantum GR, for which explicit checks and computations can be performed (see the Conclusion for 
more details). 

\vspace{2in}

%========================================================================================================
%========================================================================================================
\section{Configuration spaces of shapes}
%========================================================================================================
%========================================================================================================

This Section brings in selected material from the mathematical literature by Kendall et al. 
\cite{Kendallbook, KendallandKendall, Kendall84} for subsequent application to mechanics in Sec 3.  

\mbox{ }  

\noindent{\bf Notation} I use $\langle \mbox{ }  \rangle$ to denote `space of'. 
I add a prefix to this to indicate what kind of structure each space is, e.g. g for group, t for 
topological space, m for metric space or r for Riemannian geometry.  
The additional structures for each of these appear after a semicolon inside the angled brackets.  
$\stackrel{h}{=}$ denotes equal up to homeomorphism and $\stackrel{i}{=}$ denotes equal up to isometry.
$\mbox{Tr}(\d)$ are the d-dimensional translations. 
$\mbox{Rot}(\d)$ are the d-dimensional rotations.
$\mbox{Dil}(\d)$ are the d-dimensional dilations.  
$\mbox{Ref}(\d)$ are the d-dimensional reflections.  
$\mbox{Eucl}(\d)$ is the Euclidean group consisting of d-dimensional translations, rotations and reflections.  
$\mbox{SEucl}(\d)$ is the special Euclidean group consisting of d-dimensional translations and rotations. 
$\mbox{Sim}(\d)$ is the similarity group consisting of d-dimensional translations, rotations, dilations and reflections.  
$\mbox{SSim}(\d)$ is the special similarity group consisting of d-dimensional translations, rotations and dilations.

dim($\fQ(\N,\d)$) = $\N$d.  This space is $\mathbb{R}^{\sN\sd}$, which carries `+' and `$\cdot$' operations.  
dim(Tr(d)) = d.  
Tr(d) is $\mbox{}_{\sg}\langle\mathbb{R}^{\sd}; + \rangle$.  
dim(Rot(d)) = d(d -- 1)/2.  
Rot(d) is $SO$(d) (and there are `no rotations' in dimension 1 -- it is a discontinuous $\mathbb{Z}_2$ then).
dim(Dil(d)) = 1, while Dil(d) = Dil = $\mbox{}_{\sg}\langle\mathbb{R}^+, \cdot \rangle$, independently of d.  
Ref(d) is discrete. 
It is $\mathbb{Z}_2$.  
SEucl(d) = Tr(d) $\times$ Rot(d), SSim(d) = Tr(d) $\times$ Rot(d) $\times$ Dil = SEucl $\times$ Dil.  
Eucl(d) = SEucl(d) $\times \mathbb{Z}_2$.  
Sim(d) = SSim(d) $\times \mathbb{Z}_2$.  
dim(Eucl(d)) = d(d + 1)/2 and dim(Sim(d)) = d(d + 1)/2 + 1.

\mbox{ }

\noindent{\bf Definition 1}
{\it Relative space} $\fR(\N, \d)$ is the quotient space $\fQ(\N,\d)/\mbox{Tr}(\d)$ and 
{\it proper relative space} is $\fr(\N, \d) = \fQ(\N,\d)/\mbox{Tr}(\d) \times \mathbb{Z}_2$.  
{\it Relational space} is ${\cal R}(\N, \d) = \fQ(\N,\d)/\mbox{SEucl}(\d)$ and 
{\it proper relational space} is $\scR(\N, \d) = \fQ(\N,\d)/\mbox{Eucl}(\d)$.  
{\it Preshape space} is $\fP(\N, \d) = \fQ(\N,\d)/\mbox{Tr}(\d) \times \mbox{Dil}$ and 
{\it proper preshape space} is $\fp(\N, \d) = \fQ(\N,\d)/\mbox{Tr}(\d) \times \mbox{Dil} \times \mathbb{Z}_2$.  
{\it Shape space} is $\fS(\N, \d) = \fQ(\N,\d)\ 0/\mbox{SSim}(\d)$ and 
{\it proper shape space} is $\fs(\N, \d) = \fQ(\N,\d)\ 0/\mbox{Sim}(\d) \times \mathbb{Z}_2$.  

\mbox{ }

\noindent We will really consider the spaces of Definition 1 as augmented to be  
normed spaces, metric spaces, topological spaces, and, where possible, Riemannian geometries.  
Quotienting out the translations is so simple that it does not really matter at which 
stage this is done.  
Doing so amounts to centering the material points about a particular point.  
One can arrive at (proper) shape space either via (proper) relational space or via (proper) preshape 
space.

dim($\fR(\N, \d)$) = Nd, dim(${\cal R}(\N, \d)$) = d\{2N + 1 -- d\}/2, 
dim($\fP(\N, \d)$) = nd -- 1, D($\N$, d) $\equiv$ dim($\fS(\N, \d)$) = d\{2N + 1 -- d\}/2 -- 1, with the 
notable special cases dim(R(1, d)) = dim(${\cal R}$(1, d)) = dim($\fP$(1, d)) = dim($\fS$(1, d)) = 0, 
dim($\fP$(2, 1)) = dim($\fS$(2, 1)) = 0 and dim(${\cal R}$(2, 3)) = dim($\fS$(2, 2)) = 0 and 
dim($\fR$(2, 1)) = dim(${\cal R}$(2, 2)) = dim($\fP$(2, 2)) = 1.
See also Table 1 i) to iv).  

I use the following index types. $\alpha$ for spatial coordinates 1 to d.  

\scriptsize

\noindent
\begin{tabbing}
\underline{Real coord labels}          \hspace{1.5in}                 \= 
\underline{Complex coord labels}       \hspace{1.45in}                \= 
\underline{Configuration space coords}                                \= \\
${\cal A}$ = 1 to $\N$ = \n + 1  particle labels                      \>    
                                                                      \>
                                                                      \> \\
$A$ = 1 to N relative particle position labels                        \>
{\tt A} = 1 to \n \mbox{ } complex relative space labels in 2d        \>
A = 1 to Nd relative space coords                                     \> \\
                                                                      \>
                                                                      \> 
$\Delta$ = 1 to Nd -- 1  preshape space coords                        \> \\
a = 1 to N -- 1 radial shape coord labels in 2d                       \>
{\tt a} = 1 to N -- 1 complex shape coord labels in 2d,               \>    
a = 1 to 2$\{\sn - 1\}$ real shape coord labels in 2d,                \> \\
                                                                      \>  
comprising both the a and the $\tilde{\mbox{a}}$                      \>    
comprising both the a and the $\tilde{\mbox{a}}$                      \> \\
$\tilde{\mbox{a}}$ = 1 to N -- 1 angular shape coord labels           \>
                                                                      \>
                                                                      \> \\
\end{tabbing}

\normalsize

\noindent{\bf Definition 2} $Q_{\sN, \sd}$ is the nontrivial {\it quotient map}: 
$\fS(\N, \d) \longrightarrow \fP(\N, \d)$.  
Below I often abbreviate this to $Q$.  

\mbox{ }

\noindent{\bf Lemma 1 (real representations)} 
$\fQ(\N, \d)$ can be represented by $\bq = \langle \q_{\cal A} \mbox{ } {\cal A} = 1 \mbox{ to } \N 
\rangle$, the particle position coordinates.   
$\fR(\N, \d)$ can be represented by $\br = \langle \r_A \mbox{ } A = 1 \mbox{ to } \n \rangle$, a set of independent relative 
coordinates, e.g. 1) an independent set chosen from among the relative particle positions $\r_{\cal AB} = 
\q_{\cal A} - \q_{\cal B}$ or 2) the Jacobi coordinates $\R_A$.  
$\fP(\N, \d)$ can be represented by $\bar{\br} = \langle \bar{r}_i i = 1 \mbox{ to } \n \rangle$ obtained from these by 
normalization (so that these have 1 degree of freedom less).

%=======================================================================================================
\subsection{Topological space, metric space and topological manifold structures}
%=======================================================================================================

Note that $\fA(\d) \stackrel{h}{=} \mathbb{R}^{\sd}$, $\fQ(\N, \d) \stackrel{h}{=} 
\mathbb{R}^{\sN\sd}$, $\fR(\N, \d) \stackrel{h}{=} \mathbb{R}^{\sn\sd}$.  

\mbox{ } 

\noindent{\bf Theorem 1} $\fP(\N, \d) \stackrel{h}{=} \mathbb{S}^{\sn\sd - 1}$.

\noindent\underline{Proof} Elementary, given how $\fP(\N, \d)$ is defined.

\mbox{ }

\noindent{\bf Lemma 2} 
The \{Nd --1\}-sphere may be coordinatized by ${\cal X}_{\Delta} = x_{A\alpha}/x_{11}$ 
with the 1 among these struck out: the {\it Beltrami coordinates}.  

\mbox{ } 

\noindent{\bf Structure 1} The $\mathbb{R}^{\sd}$ inner product serves to have a notion of 
`localized in space', which survives in some form for all the configuration spaces considered.  
This is useful as regards discussing observable configurations.  

\mbox{ }  

\noindent{\bf Structure 2} One also has have a notion of localized in configuration space -- i.e. of which 
configurations look alike (this is important as in physics one does not know precisely what 
configuration one has).  
For this one has available the possibly-weighted $\mathbb{R}^{\sN\sd}$ norm  $||\mbox{ }||_{\sw}$.   
One could use the $\mathbb{R}^{\sn\sd}$ norm in $\fP(\N, \d)$ too, as a chordal distance, but there are also 
intrinsic distances that could be used thereupon (based on angular separations in $\mathbb{S}^{\sn\sd - 1}$).  

\mbox{ } 

\noindent{\bf Structure 3}   
$ _m\langle \fQ(\N, \d), ||\mbox{ }||_{\sw}\rangle$ and 
$ _m\langle \fR(\N, \d), ||\mbox{ }||_{\sw}\rangle$ are appropriate as metric spaces to work with.  
Then one inherits $ _m\langle \fP(\N, \d), D_{\mbox{\scriptsize chordal}}\rangle$, which is 
topologically equivalent to $ _m\langle \fQ(\N, \d), D_{\mbox{\scriptsize great circle}}\rangle$, see  
\cite{Kendallbook} p13 [$D_{\mbox{\scriptsize chordal}}$ and $D_{\mbox{\scriptsize great circle}}$ being related  
by $D_{\mbox{\scriptsize chordal}} = \mbox{sin}(D_{\mbox{\scriptsize great circle}})$, see 
\cite{Kendallbook} p205].  
Then $D_{\mbox{\scriptsize great circle}}(\bar{A}, \bar{B}) = \mbox{Arccos}(\bar{A}\cdot\bar{B}).$  
This carries over to shape space: one has the metric space 
$ _m\langle \fS(\N, \d), D\rangle$, for the quotient metric
\beq 
D(Q(\bar{A}, \bar{B}) = \mbox{ } \stackrel{\mbox{min}}{T \in SO(\d)} 
D_{\mbox{\scriptsize great circle}}(\bar{A}, T\bar{B}) 
 = \mbox{ } \stackrel{\mbox{min}}{T \in SO(\d)} \mbox{arccos}(\bar{A} \cdot T\bar{B}) \mbox{ } .  
\label{KenMin}
\eeq

\noindent{\bf Structure 4}  Appropriate topological spaces to work with are 
$ _t\langle \fP(\N, \d); \tau_{\sfP}\rangle$ for $\tau_{\sfP}$ the set of open sets (obeying topological space 
axioms) determined by $D_{\mbox{\scriptsize chordal}}$ or $D_{\mbox{\scriptsize great circle}}$ and 
$ _t\langle \fS(\N, \d); \tau_{\sfS}\rangle$ e.g. obtained from the preceding as the quotient topology 
corresponding to the map Q, or, equivalently, as the set of open sets determined by 
$D_{\mbox{\scriptsize chordal}}$ or $D_{\mbox{\scriptsize great circle}}$.  
As an available resource, untapped in the present Paper, p12 \cite{Kendallbook} provides some theorems 
for shape spaces at the level of topological spaces.  

\mbox{ } 

\noindent{\bf Lemma 3 (complex representations).} 
i) In 2d, relative configurations can be represented by N complex numbers 
$\langle z_{\mbox{\scriptsize \tt A}} \mbox{ } \mbox{\tt A} = 1 \mbox{ to } \n\rangle$ -- the 
{\it homogeneous coordinates}.  

\noindent ii) Assuming that not all of these are simultaneously 0 (i.e. excluding the maximal collision), 
2d shapes can be represented by N -- 1 independent complex ratios, the so-called {\it inhomogeneous 
coordinates} $\langle{\cal Z}_{\mbox{\scriptsize \tt a}} \mbox{ } \mbox{\tt a} = 1 \mbox{ to } $N$ - 1 
\rangle$ [This is established by dividing the complex numbers in i) by a particular 
$z_{\mbox{\scriptsize \tt A}}$ and ignoring the 1 among the new string of complex numbers.]

\mbox{ } 

\noindent{\bf Theorem 2 (shape spaces as topological surfaces)}.

\noindent i) $\fS(\N, 1) \stackrel{h}{=} \mathbb{S}^{\sN - 2}$

\noindent ii) $\fS(\N, 2) \stackrel{h}{=} \mathbb{CP}^{\sN - 2}$

\noindent iii) $\fS(\d + 1, \d) \stackrel{h}{=} \mathbb{S}^{\sC(\sd)}$ for $\mbox{C}(\d) = D(\d + 1, \d)$

\noindent iv)  $\fS(\mA, \mB) \stackrel{h}{=} \mathbb{RP}^{\sC(\sB)}$ for $\mA \leq \mB$.

\noindent(cf Table 1 v)).

\noindent\underline{Proof}
i) is a trivial corollary of Theorem 1. 

\noindent ii) Use Lemma 3.ii) (which is a standard presentation of $\mathbb{CP}^{\sN - 2}$). 
Then take out $e^{i\alpha}$ [which amounts to taking out the $SO(2)$ rotations].  

\noindent iii) is Casson's theorem (see \cite{Kendallbook} p20-22).

\noindent iv) is a simple corollary of Casson's theorem, which follows as one is now allowed to make 
reflections via higher-d rotations, and $\mathbb{RP}^{\sk} = \mathbb{S}^{\sk}/\mathbb{Z}_2$ $\Box$.  

\mbox{ } 

\noindent Note that there is indeed agreement on the various overlaps of the above. C(2) = 0 so $\mathbb{S}^{\sC(2)} = 
\mathbb{S}^{2 - 2}$.  

\noindent
$\mathbb{RP}^{\sC(2)} \equiv \mathbb{RP}^0 \stackrel{h}{=} \mathbb{CP}^0 \equiv \mathbb{CP}^{2 - 2}$. 
$\mathbb{CP}^1 \stackrel{h}{=} \mathbb{S}^2$ is a well-known geometrical result.
Also note that the missing triangle in Table 1 v) are likely new in this context rather than known from 
elsewhere in mathematics.  
Finally note that at the topological level, 1d and 2d shape spaces are in terms of standard spaces. 
At the topological level, the 3d spaces are $\mathbb{RP}^2$, $\mathbb{S}^5$, and then all new spaces, 
while the 4d spaces are $\mathbb{RP}^2$, $\mathbb{RP}^5$, $\mathbb{S}^9$ and then all new spaces.  
See Ch. 2-5 of \cite{Kendallbook} for a partial characterization of these spaces at the topological level.  

\mbox{ } 

\noindent{\bf Corollary 1} i)  $\fs(\N, 1) \stackrel{h}{=} \mathbb{RP}^{\sN - 2}$.

\noindent ii) $\fs(\N, 2) \stackrel{h}{=} \mathbb{CP}^{\sN - 2}/\mathbb{Z}_2$.

\noindent iii) $\fs(\mbox{A}, \mbox{B}) \stackrel{h}{=} \mathbb{RP}^{\sC(\sB)}$ 
for $\mbox{A} \leq \mbox{B} + 1$

\noindent [See also Table 1.vi)].

\noindent\underline{Proof} Use Theorem 2 and note that quotienting out twice is the same as quotienting out once 
for the last part $\Box$.  

\mbox{ } 

\noindent{\bf Corollary 2} $\fR(\N, \d) \stackrel{h}{=} \fP(\N,\d) \times [0, \infty)$ and 
${\cal R}(\N, \d) \stackrel{h}{=} \fS(\N, \d) \times [0, \infty)$, though these do not exclude the 
possibility of pathologies at the origin. 

\noindent\underline{Proof} Build up an `onion'.  

\mbox{ } 

\noindent Note that, because of pathologies at the origin, relational space is in some ways a less 
advantageous intermediate to study than preshape space.

%=======================================================================================================
\subsection{Riemannian metric structure}
%=======================================================================================================

I begin with two simple results.  

\mbox{ } 

\noindent{\bf Theorem 3} $\fP(\N, \d) \stackrel{i}{=} \mbox{}_{\sr}\langle \mathbb{S}^{\sn\sd - 1}, 
g_{\Gamma\Delta}^{(\mbox{\scriptsize sphere, 1/2})}\rangle$. [`Sphere, 1/2' means the standard 
spherical metric of curvature 1/2.]

\noindent\underline{Proof} \fP(\N, \d) is described by $\sum_{\Delta = 1}^{\sn\d}\bar{r}_{\Delta}^2$ = constant  
(normalization condition, which is clearly the $\mathbb{S}^{\sn\d - 1}$  sphere embedded in the usual way in $\mathbb{R}^{\sn\d}$ $\Box$.  

\mbox{ } 

\noindent{\bf Theorem 4} In the Beltrami coordinates of Lemma 2, the standard metric on the 
\{Nd--1\}-sphere has line element 
\beq
\d s^2 = M_{\Gamma\Delta}\d{\cal X}^{\Gamma}\d{\cal X}^{\Delta} \equiv  
\frac{\{1 + ||\mbox{\boldmath${\cal X}$}||^2 \} ||\mbox{\boldmath$\d{{\cal X}}$}||^2 - 
(\mbox{\boldmath${\cal X}$}, \mbox{\boldmath$\d{{\cal X}}$})^2 }{\{1 + ||\mbox{\boldmath${\cal X}$}||^2 \}^2} \mbox{ } .  
\label{Belt}
\eeq

\noindent I next consider the situation for shape space.  

\mbox{ } 

\noindent{\bf Structure 5} Associated with the nontrivial quotient map Q are the orbits 
Orb($\bar{X}$) = $Q^{-1}(Q(\bar{X})) = \langle T\bar{X} | T \in SO(\d) \rangle$ and the stabilizers 
Stab($\bar{X}$) = $\langle T \in SO(\d) | T\bar{X} = \bar{X} \rangle$.  
These can furthermore be thought of as fibres and isotropy groups.  
The orbits or fibres are, for $\bar{X}$ of rank e, 
$Q(\bar{X}) =\left\{ \stackrel{\mbox{$SO$(\d), e $\geq \d$ + 1}}
                             {\mbox{Stie(d, e), e $< \d$ + 1}}\right.$ 

\noindent for Stie(d, e) = $SO$(d)/$SO$(d -- e) the Stiefel manifolds \cite{Hatcher} of orthonormal e frames in $\mathbb{R}^{\sd}$. 
As an available resource untapped in this Paper, 
\cite{Kendallbook} provides a number of topological space results about the fibres.  

\mbox{ } 

\noindent{\bf Definition 3}  Let ${\cal D}_e(\N, \d)$ be the subset of $\fP(\N, \d)$ corresponding to 
rank $\leq$ e and let ${\cal D}_{\se}^{\sc}(\N,\d)$ be its complement.  
I will often drop the $(\N, \d)$ from the notation.

\mbox{ } 

\noindent Note that restricting attention to shape spaces on or above the Casson diagonal,  
${\cal D}_{\sN - 2}$ is the set with nontrivial isotropy groups.  

\mbox{ } 

\noindent{\bf Definition 4} $Q({\cal D}_{\sN - 2})$ is the {\it singularity set} of $\fS(\N, \d)$, 
while $Q({\cal D}_{\sN - 2}^{\sc})$ is the {\it nonsingular part} of $\fS(\N, \d)$.  

\mbox{ } 

\noindent{\bf Theorem 5} On $Q({\cal D}_{\sN - 2}^{\sc})$ there is a unique Riemannian metric compatible 
with the differential structure and with respect to which Q is particularly well behaved is inherited 
from $\fP(\N, \d)$.

\noindent\underline{Proof} 1) by Riemannian submersion \cite{Neill}.  
2) Alternatively, from first principles according to the steps below up to and including Structure 7 $\Box$.  

\mbox{ } 

\noindent{\bf Lemma 4} On $\fP(\N, \d)$ the geodesics are great circles.  

\noindent\underline{Proof} By Theorem 3, $\fP(\N, \d) \stackrel{i}{=}  \mathbb{S}^{\sn\sd - 1}$,  
and then it is well-known that the geodesics of spheres are great circles $\Box$.  

\mbox{ } 

\noindent Note that the geodesic joining $\bar{X}$ and $\bar{Z}$ takes the form 
\beq
\Gamma_{\bar{Z}}(s) = \bar{X}\mbox{cos}s + \bar{Z}\mbox{sin}s
\label{GeoForm}
\eeq 
parametrized in terms of geodesic distance, for $0 \leq s \leq \pi$.
%
%There is a geodesic length formula available if required.
%
%
Also note that the tangent vector to the geodesic is 
$\left.\frac{\d \Gamma_{\bar{Z}}(s)}{\d s}\right|_{s = 0} = \bar{Z}$.  

\mbox{ } 

\noindent{\bf Structure 6} i) The {\it exponential map} is $T_{\bar{X}}(\fP(\N, \d)) \longrightarrow 
\fP(\N, \d)$ $\bar{Z} \longmapsto \Gamma_{\frac{\bar{Z}}{||Z||}}(||Z||)$.
It restricts to a diffeomorphism of $\langle \bar{Z} \in T_{\bar{X}}(\fP(\N, \d))  | \mbox{ } ||\bar{Z}|| < \pi 
\rangle$ onto $\langle \fP(\N,\d)/\mbox{antipode of } \bar{X} \rangle$.  

$\hat{\Gamma}_{A}$ is a curve in $SO$(d) starting from $\mathbb{I}$ so that the tangent vector at $s = 0$, 
$\left.\frac{\d\hat{\Gamma}_{A}}{\d s}\right|_{s = 0} = A$ is tangent to $SO$(d) at $\mathbb{I}$. 
$\mbox{exp}(sA)$ is in $SO$(d) iff $\mbox{exp}(sA)\mbox{exp}(sA)^T = \mathbb{I}$ iff 
exp$(s\{A + A^{\sT}\}) = \mathbb{I}$  iff $A + A^{\sT} = 0$. 
So any skew-symmetric matrix $\bar{A}$ represents a vector tangent to $SO$(d) at $\mathbb{I}$.  
As the space of d $\times$ d skew symmetric matrices has dimension d(d -- 1)/2 = dim($SO$(d)), that's the 
entire tangent space to $SO$(d) at $\mathbb{I}$.  

\noindent ii) as exp$(s\bar{A})$ lies in $SO$(d) whenever $A^{\sT} = -A$, $\hat{\gamma}_A(s) = 
\mbox{exp}(sA) \bar{X}$ lies in the fibre orbit through $\bar{X}$.  

\noindent{\bf Definition 5} The subspace of tangent vectors 
$\left.\frac{\d\hat{\gamma}_A(s)}{\d s}\right|_{s = 0} = A\bar{X}$ to such curves at $\bar{X}$ is the 
{\it vertical} tangent subspace at $\bar{X}$, $V_{\bar{X}} = \langle A\bar{X}| A^{\sT} = A\rangle$.
Its orthogonal complement $\fH_{\bar{X}}$ i.e. such that $T_{\bar{X}}(\fP(\N, \d)) = 
\fV_{\bar{X}}(\N, \d) \oplus \fH_{\bar{X}}(\N, \d)$ is the {\it horizontal} tangent subspace at $\bar{X}$. 
%
%It's a principal bundle.

\mbox{ } 

\noindent Note that for $Q(\bar{X})$ nonsingular, i.e. 
$\bar{X} \not{\in\hspace{0.03in}} {\cal D}_{\sd - 2}$, $\fV_{\bar{X}}(\N, \d)$ is isomorphic to $SO$(d) at 
$\mathbb{I}$, but at a 
singular point $A$ is tangent to the isotropy subgroup, so $\fV_{\bar{X}}(\N, \d)$ is isomorphic to Stie(d, e) 
at $\mathbb{I}$.
%
%Have available a formula for H's.

\mbox{ } 

\noindent{\bf Proposition 1} 

\noindent i) If a geodesic in $\fP(\N, \d)$ starts out in a horizontal direction, 
then its tangent vectors remain horizontal along it.    

\noindent ii) Distance-parametrization and horizontality are preserved under $SO$(d).

\noindent\underline{Proof} i) by definition, the geodesic $\Gamma_{\bar{Z}}(s)$ is horizontal at 
$s = 0$ iff $\bar{X}\bar{Z}^{\sT} = \bar{Z}\bar{X}^{\sT}$.  
Then $\forall \mbox{ } s$, $\Gamma_{\bar{Z}}\left\{\frac{\d \Gamma_{\bar Z}}{\d s}\right\} = 
\frac{\d \Gamma_{\bar{Z}}}{\d s}\Gamma_{\bar{Z}}(s)^{\sT}$ via the explicit formula (\ref{GeoForm}) 
for the geodesics and trivial algebra.  
Thus each tangent vector $\frac{\d \Gamma_{\bar{Z}}}{\d s}$ complies with the definition of horizontal 
at $\Gamma_{\bar{Z}}(s)$.  

\noindent ii) $T\Gamma_{\bar{Z}}(s) = T\bar{X}\mbox{cos}(s) + T\bar{Z}\mbox{sin}(s)$ is a 
distance-preserving geodesic by (\ref{GeoForm}), and $\mbox{tr}(T\bar{X}\{T\bar{Z}\}^{\sT}) = 
\mbox{tr}(T\bar{X}\bar{Z}^{\sT}T^{\sT}) = \mbox{tr}(T^{\sT}T\bar{X}\bar{Z}^{\sT}) = 
\mbox{tr}(\bar{X}\bar{Z}^{\sT})$ by the cyclic identity and $T$ orthogonal.
Next, $T\bar{X}\{T\bar{Z}\}^{\sT} = T\bar{X}\bar{Z}^{\sT}T^{\sT} = T\bar{Z}\bar{X}^{\sT}T^{\sT} = 
T\bar{Z}\{T\bar{X}\}^{\sT}$ (with the third step using horizontality of the untransformed geodesic), 
which reads overall that the transformed geodesic is then also horizontal $\Box$.

\mbox{ } 

\noindent{\bf (Riemannian) Structure 7} i) Thus the exp function restricted to $\fH_{\bar{X}}$, 
exp$|_{\sfH_{\bar{X}}}$, maps 
$\langle$ vectors of length $< \pi$ $\rangle$ onto the submanifold $\mathbb{H}_{\bar{X}}$ of 
$\fS(\N, \d)$ 
defined by $\mathbb{H}_{\bar{X}} =  \langle s \in \fS(\N, \d) |$ all tangent vectors at $\bar{X}$ are horizontal
$\rangle$.  

\noindent ii) Since the tangent spaces to the fibre and to $\mathbb{H}_{\bar{X}}$ are clearly perpendicular, 
there is a neighbourhood $U_{\bar{X}}$ such that $\forall \mbox{ } Y \in U_{\bar{X}}$ the tangent spaces 
to the fibre and to $\mathbb{H}_{\bar{X}}$ remain transverse.  
Thus the fibre at Y meets $U_{\bar{X}}$ only at Y. 
Thus, one has established that given X outside ${\cal D}_{\d - 2}$, through each point $T\bar{X}$ of the 
fibre at $\bar{X}$ there is a submanifold $U_{\bar{X}}$ traced out by local horizontal geodesics through 
$T\bar{X}$ such that  

\noindent a) $U_{T_1\bar{X}}$ and $U_{T_2\bar{X}}$ are disjoint if $T_1 \neq T_2$.

\noindent b) Each submanifold $U_{T\bar{X}}$ is mapped by the quotient mapping Q bijectively and thus  
homeomorphically with respect to the quotient topology onto a neighbourhood of $Q(\bar{X}) \in \fS(\N, \d)$.  

\noindent c) The action of each $S \in$ $SO$(d) restricts to a diffeomorphism of $U_{T\bar{X}}$ that also preserves 
the Riemannian metric, i.e. it maps geodesics to geodesics of the same length and its derivative maps 
horizontal tangent vectors at $T\bar{X}$ to horizontal tangent vectors of the same length at $ST\bar{X}$.  
Thus one can use $U_{\bar{X}}$, $Q|_{U_{\bar{X}}}$ to determine a differential structure on the 
nonsingular part of shape space: $Q({\cal D}_{\d - 2})$, since for any other choice 
($U_{T\bar{X}}$, $Q|_{U_{T\bar{X}}}$) the composition $(Q|_{U_{T\bar{X}}})^{-1} \circ Q|_{U_{\bar{X}}}$ 
is just the diffeomorphism $T|_{U_{\bar{X}}}$.  

\noindent iii) The above ensures independence of which point on the fibre is used. 
Thus we have a Riemannian metric on the nonsingular part of shape space.  

\mbox{ } 

\noindent Note that this metric is naturally induced from $\fP(\N, \d) = \mathbb{S}^{\sn\sd  - 1}$.   
%
%[Viewed in this way, it is the {\it Hopf metric}, but that is not relevant to this work, nor a very well known notion, it would seem].  
%
It has been defined such that 

\noindent $Q: \fH_{\bar{X}}(\N,\d) \longrightarrow \fP(\N, \d) \stackrel{i}{\longrightarrow} 
T_{Q(\bar{X})}(\fS(\N, \d))$, which is a Riemannian submersion.  

\mbox{ } 

\noindent{\bf Definition 6} A {\it geodesic} in $\fS(\N, \d)$ is the image of any horizontal 
geodesic in $\fP(\N, \d)$.  

\mbox{ } 

\noindent Note that this permits geodesics to pass between strata.  
Thus geodesics can be extended beyond the nonsingular part of the space, and this serves to extend the 
above Riemannian structure (see \cite{Kendallbook} for these results, which not used in the present 
paper).
%
%Suppressed second part of remark: Minimal geodesics for $\fS(\N, \d)$ are not necessarily unique.

\mbox{ } 

\noindent{\bf Proposition 2} The geodesics of (\ref{GeoForm}), Lemma 4
(or the associated Riemannian metric of Structure 7) provide the same metric distance D as Structure 3.  

\noindent\underline{Proof} By definition, the geodesic between two shapes $Q(\bar{X})$ and $Q(\bar{Y})$ 
is the image of a horizontal geodesic $\Gamma$ from $\bar{X}$ to some point T$\bar{Y}$ in fibre 
$Q(\bar{Y})$. 
Since $\Gamma$ meets the fibres orthogonally at these points [$\Gamma$ being horizontal 
and the fibres being vertical], so the induced distance that follows from the geodesics/associated 
Riemannian metric is indeed 

\noindent $\stackrel{\mbox{min}}{T \in SO(\d)}D(\bar{X}, T\bar{Y}) = 
\mbox{arccos} \stackrel{\mbox{max}}{T \in SO(\d)} \mbox{tr}(T\bar{Y}\bar{X}^{\sT}) \equiv 
D(Q(\bar{X}), Q(\bar{Y})$ $\Box$.    

\mbox{ } 

\noindent Note that what one has constructed thus is a Riemannian structure on 
$Q({\cal D}_{\sd - 2}^{\sc})$.
In general, one would have to worry about the geometry on ${\cal D}_{\sd - 2}$ -- the name 
`singularity set' does indeed carry curvature singularity connotations.  
But this paper circumvents that by considering 
only dimension 1 and dimension 2, for which the singularity set is empty.    
Thus for these cases, what one has constructed above is a Riemannian structure everywhere on shape 
space (and one can then show by computation thereupon that there are no curvature singularities 
within these shape spaces).

%========================================================================================================
\subsection{1d shape spaces}
%========================================================================================================

\noindent{\bf Lemma 5} $\fS(\N, 1) = \fP(\N, 1)$, ${\cal R}(\N, 1) = R(\N, 1)$ 
(both homeomorphically and isometrically).  

\mbox{ } 

\noindent{\bf Corollary 3} $\fS(\N, 1) \stackrel{h}{=} \fP(\N, 1) \stackrel{h}{=} \mathbb{S}^{\sN - 2}$.  

\noindent\underline{Proof} $\fS(\N, 1) = \fP(\N, 1)$ by the rotations being trivial in 1d.  
Then use Theorem 1:  $\fP(\N, 1) \stackrel{h}{=} \mathbb{S}^{\sn\sd - 1} \equiv \mathbb{S}^{\sN - 2}$.   

\mbox{ } 

\noindent Note that, from the triviality of the rotations involved, nothing needs to be induced from the sphere, 
nor is any minimization required.  
The singularity set is empty.

\mbox{ } 

\noindent{\bf Corollary 4} $\fS(\N, 1) \stackrel{i}{=} \fP(\N, 1) \stackrel{i}{=} \mbox{}_{\sr}\langle 
\mathbb{S}^{\sN - 2}, g_{\Gamma\Delta}^{(\mbox{\scriptsize sphere, 1/2})}\rangle$.  

\noindent\underline{Proof} $\fS(\N, 1) = \fP(\N, 1)$ by rotations being trivial in 1d.  
Then use Theorem 3: $\fP(\N, 1) \stackrel{i}{=} \mathbb{S}^{\sn\sd - 1} \equiv \mathbb{S}^{\sN - 2}$ 
$\Box$.  

\mbox{ } 

\noindent{\bf Corollary 5} The metric on 1-d shape space is the appropriate subcase of (\ref{Belt}).

%========================================================================================================
\subsection{2d shape spaces}
%========================================================================================================

In 2d rotations are simpler than in higher d while being nontrivial.  
It is this that lies behind Lemma 3's straightforward complex representations for $\fP(\N, 2)$ and 
$\fS(\N, 2)$.  
The latter representation, $\langle {\cal Z}_{\mbox{\scriptsize \tt a}} \mbox{ } \mbox{\tt a} = 1 
\mbox{ to } \n - 1 \rangle$ has two manifest symmetries: $\mathbb{Z}_2$ conjugation and 
$U(\N - 1)/U(1)$ permutations of coordinates.  
Each of these commutes with the $SO(2)$ action.

Other simplifications in 2d are that $S(\N, 2) \stackrel{h}{=} \mathbb{CP}^{\sN - 2}$ [Theorem 2 ii)], and note that
the minimization in the quotient metric (in the metric space sense) of Structure 3 may be carried out 
explicitly in 2d by use of the complex representation as follows.   

\mbox{ } 

\noindent{\bf Proposition 3} $\mbox{cos}D(Q(\bz), Q(\bw)) = 
\frac{|(\bz \cdot \bw)_{\tC}|}{||\bz||_{\tC}||\bw||_{\tC}}$ 
for $(\mbox{ }\cdot\mbox{ })_{\tC}$ the $\mathbb{C}^n$ inner product and $||\mbox{ }||_{\sC}$ the 
corresponding norm.  

\noindent\underline{Proof} By the definition of $D$, the left-hand side is 
$\stackrel{\mbox{max}}{\alpha \in [0, 2\pi)}\mbox{tr}(|\bz|^{-1}|\bw|^{-1}\mbox{e}^{i\alpha}\bz\bw^{\sT})$, 
the numerator of which contains 
$$
\stackrel{\left(\mbox{Re($\bz$)} \mbox{ } \mbox{Im($\bz$)}\right)}{\mbox{ }}
\mbox{\Huge $($}
\stackrel{\mbox{ }  \mbox{cos$\alpha$} \mbox{ } \mbox{sin$\alpha$}}
         {\mbox{--} \mbox{sin$\alpha$} \mbox{ } \mbox{cos$\alpha$}}
\mbox{\Huge $)$}
\mbox{\Huge $($}
\stackrel{\mbox{Re($\bw$)}}{\mbox{Im($\bw$)}}
\mbox{\Huge $)$}
= A\mbox{cos}\alpha + B\mbox{sin}\alpha \mbox{ } ,
$$
where $A = \mbox{Re}(\bw \cdot \bz)_{\sC}$ and $B = \mbox{Im}(\bw \cdot \bz)_{\sC}$.  
The maximum condition which follows from this is then tan$\alpha = B/A$, for which the maximum value is 
$\sqrt{A^2 +  B^2}/||\bz||_{\sC}||\bw||_{\sC}$ $\Box$.    

\mbox{ } 

\noindent{\bf Theorem 6} The corresponding Riemannian line element is 
\beq
\d D^2 = \frac{||\bz||_{\sC}^2 ||\d \bz||_{\sC}^2 - |(\bz \cdot \d \bz)_{\sC}|^2}
              {||\bz||_{\sC}^4}
       = \frac{\{1 + ||\mbox{\boldmath${\cal Z}$}||_{\sC}^2\} ||\d \mbox{\boldmath${\cal Z}$}||_{\sC}^2 - 
                 |(\mbox{\boldmath${\cal Z}$} \cdot \d \mbox{\boldmath${\cal Z}$})_{\sC}|^2}
              {\{1 + ||\mbox{\boldmath${\cal Z}$}||_{\sC}^2\}^2}  \mbox{ } .  
\label{Easter}
\eeq
\noindent\underline{Proof}  Consider $\bw = \bz + \delta \bz$.  
Then 
\beq
\delta D^2 = \mbox{sin}^2\delta D + O(\delta D^4) = 
1 - \mbox{cos}^2 D(Q(\bz), Q(\bz + \delta \bz) + O(\delta D^4) 
\eeq
then use the Proposition 3, linearity, the binomial expansion and take 
the limit as $\delta \bz \longrightarrow 0$ to get the first form.  
Then divide top and bottom by $||z_i||_{\sC}^4$ and use the definition of ${\cal Z}_{\mbox{\scriptsize \tt a}}$ to get 
the second form $\Box$.  

\mbox{ } 

\noindent This line element (which indeed is Riemannian, its positive-definiteness following from the 
Schwarz inequality) is the classical Fubini--Study \cite{FS} line element on $\mathbb{CP}^{\sN - 2}$, 
which is the natural line element thereupon, such that its constant curvature is 4.

Thus the following has been proven.  

\noindent{\bf Corollary 6} $S(\N, 2) \stackrel{i}{=} \mbox{}_{\sr}\langle \mathbb{CP}^{\sN - 2}; 
g^{(\mbox{\scriptsize Fubini--Sudy, 4})}_{\sa\sb} \rangle$.   

\mbox{ } 

\noindent{\bf Corollary 7} $S(3, 2) \stackrel{i}{=} \mbox{}_{\sr}\langle \mathbb{S}^2; 
g^{(\mbox{\scriptsize sphere, 1/2})}_{\sa\sb} \rangle$. 

\noindent\underline{Proof} Now $|| \mbox{ } ||_{\sC} = | \mbox{ } |$, so two terms cancel in the second 
form of (\ref{Easter}), leaving
\beq
\d D^2 = \frac{\d{\cal Z}\d\overline{{\cal Z}}}{\{1 + {\cal Z}^2\}^2} = 
\frac{\d r^2 + r^2\d\theta^2}{\{1 + r^2\}^2} = 
\left\{\frac{1}{2}\right\}^2\{\d\psi^2  + \mbox{sin}^2\psi \d\theta^2\}
\eeq
by using the polar form for the complex numbers and the coordinate transformation 
$r = \mbox{tan}\frac{\psi}{2}$ $\Box$.

\mbox{ } 

\noindent Note that in 2d, the $SO(\d)$ action is free, there is no stratification, 
no complications in considering geodesics, and the natural metric on shape space is everywhere-defined 
and everywhere of finite curvature.

%========================================================================================================
\subsection{Higher-d shape spaces}
%========================================================================================================

Here, the Riemannian structure and sometimes even the topology are no longer standard, well-studied 
ones, while the singular set begins to play a prominent and obstructory role. 
\cite{Kendallbook} does however describe and provide references for a partial study of (easier subcases 
of) these shape spaces.

%========================================================================================================
\subsection{Relative and relational space counterparts}
%========================================================================================================

The maximal collision can be mathematically unpleasant.  
In this sense (proper) shape space is easier to handle than (proper) relational space.  
That is a useful guide in seeking for tractable toy models.

%========================================================================================================
%========================================================================================================
\section{Mechanics on configuration spaces of shapes}
%========================================================================================================
%========================================================================================================

For these mechanics, the incipient notion of space {\it Absolute space} $\fA(\d) = \mathbb{R}^{\sd}$ and 
the incipient {\it configuration space} $\fQ(\N, \d) = \langle\N$ labelled possibly superposed material 
points in $\mathbb{R}^{\sd}\rangle$.  
Configurational relationalism is to be directly implemented: rather than involve a nontrivial group of 
irrelevant motions, I work directly on the reduced configuration space in Sec 2, which gives me natural 
composite objects, in particular natural metrics, out of which to construct my actions.  
In each case I consider a Jacobi-type action to implement temporal relationalism through manifest 
reparametrization invariance without any variables extraneous to the configuration space,  
\beq
\fI = \int\d\lambda \fL = 2\int\d\lambda\sqrt{\fT\{\fU + \fE\}} \mbox{ } .  
\label{action2}
\eeq

%========================================================================================================
\subsection{Direct construction of a natural mechanics on preshape space}
%========================================================================================================

{\bf Theory 1} The reduced configuration space is here preshape space, which is 
$\mathbb{S}^{\sn\sd - 1}$: $|\bx|^2 = \mbox{const}$.
If one considers the natural spherical metric on this in Beltrami coordinates (\ref{Belt}), the natural 
kinetic term for a mechanics is  
\beq
\fT = \frac{1}{2}||\mbox{\boldmath$\dot{\cal X}$}||^2_{\mbox{\scriptsize\boldmath${\cal M}$}(\mbox{\scriptsize sphere})} 
\equiv \frac{1}{2}\frac{\{1 + ||\mbox{\boldmath${\cal X}$}||^2 \} ||\mbox{\boldmath$\dot{{\cal X}}$}||^2 - 
(\mbox{\boldmath${\cal X}$}, \mbox{\boldmath$\dot{{\cal X}}$})^2 }{\{1 + ||\mbox{\boldmath${\cal X}$}||^2 \}^2} \mbox{ } .  
\label{9}
\eeq
The natural potential term $\fV$ to take is a function of the ${\cal X}_{\Delta}$, i.e. a function that 
is homogeneous of degree zero in the $x_{a\alpha}$.  
The Jacobi action for a mechanics is then (\ref{action2}) with this kinetic term and potential term 
substituted in.

%========================================================================================================
\subsection{Direct construction of a natural mechanics on shape space}
%========================================================================================================

{\bf Theory 2} The reduced configiration space is here shape space.    
The natural metric kinetic term for a theory on shape space is that whose kinetic term is 
constructed using the natural shape space metric.  
In dimension 1, preshape space is shape space, so one arrives at a subcase of the theory in the 
previous Subsec.  

\mbox{ } 

\noindent In dimension 2, shape space is $\mathbb{CP}^{2\{\sn - 1\}}$, which carries the natural Fubini--Study metric 
as per Sec 2.
In the usual inhomogeneous coordinates of Lemma 3, the natural kinetic term for a mechanics is 
\beq
\fT = \frac{1}{2}
||\mbox{\boldmath$\dot{\cal Z}$}||^2_{\mbox{\scriptsize\boldmath${\cal M}$\mbox{\scriptsize(Fubini--Study)}    }     } 
\equiv \frac{1}{2}\frac{    \{1 + ||\mbox{\boldmath${\cal Z}$}||_{\sC}^2\} 
                ||\dot{\mbox{\boldmath${\cal Z}$}}||_{\sC}^2 - 
                 |(\mbox{\boldmath${\cal Z}$} \cdot \dot{\mbox{\boldmath${\cal Z}$}})_{\sC}|^2    }
            {    \{1 + ||\mbox{\boldmath${\cal Z}$}||_{\sC}^2\}^2    } \mbox{ } ,
\label{10}
\eeq  
while the natural potential term $\fV$ to take is a function of the ${\cal Z}_{\mbox{\scriptsize \tt a}}$, 
i.e. a function that is homogeneous of degree zero in the $z_{a\alpha}$.  
The Jacobi action for a mechanics is then (\ref{action2}) with this kinetic term and potential term 
substituted in.

%========================================================================================================
\subsection{How a general mechanical theory unfolds from actions of the above form}
%========================================================================================================

Consider (\ref{action2}) with
\beq
\fT = \frac{1}{2}||\mbox{\boldmath$\dot{{\cal Q}}$}||_{\mbox{\scriptsize\boldmath${\cal M}$}}^2 \mbox{ } .  
\label{Te}
\eeq
The momenta are 
\beq
{\cal P}_{\bar{p}} = \sqrt{\frac{\fE - \fV}{\fT}}{\cal M}_{\bar{p}\bar{q}}\dot{\cal Q}^{\bar{q}} \mbox{ } .  
\eeq
There is then as a primary constraint the quadratic energy constraint
\beq
\frac{1}{2}||\mbox{\boldmath${\cal P}$}||_{\mbox{\scriptsize\boldmath${\cal N}$}}^2 = 
\frac{1}{2}\left|\left|\sqrt{\frac{\fE - \fV}{\fT}}
\mbox{\boldmath${\cal M}\dot{{\cal Q}}$}\right|\right|_{\mbox{\scriptsize\boldmath${\cal N}$}}^2
= \frac{\fE - \fV}{\fT}\frac{1}{2}||\mbox{\boldmath$\dot{{\cal Q}}$}||_{\mbox{\scriptsize\boldmath${\cal MNM}$}}^2 
= \fE - \fV \mbox{ } 
\eeq
by $\mbox{\boldmath${\cal NM}$} = \mathbb{I}$. and (\ref{Te}).  
This is, as in Sec 1, a homogeneous quadratic energy constraint that is analogous to the GR Hamiltonian 
constraint and carries associated with its form the frozen formalism aspect of the problem of time.  
Variation with respect to $\underline{\cal R}_a$ gives the equation of motion of the theory, 
which is, in its geodesic equation-like velocity form,  
\beq
\nabla_t\dot{{\cal Q}}^{\bar{p}} \equiv \ddot{{\cal X}}^{\bar{p}} + 
{\Gamma^{\bar{p}}}_{\bar{q}\bar{r}}{\cal X}^{\bar{q}}{\cal X}^{\bar{r}} = 
- {\cal N}^{{\bar{p}}{\bar{q}}}{\fV}_{,{\bar{q}}} \mbox{ } , 
\eeq 
%[This requires slight modification if any cases included in paper are noninvertible].  
%
where $\Gamma$ is the metric connection.  
Or, in terms of momentum variables,   
\beq
\dot{{\cal P}}_{\bar{p}} = 
||\mbox{\boldmath${\cal P}$}||_{\mbox{\scriptsize\boldmath${\cal NM}$}_{,{\bar{p}}}
                                \mbox{\scriptsize\boldmath${\cal N}$}}^2 
- {\cal V}_{,{\bar{p}}} \mbox{ } .  
\eeq
This equation of motion does indeed propagate the energy constraint: 
\beq
\{||\mbox{\boldmath${\cal P}$}||_{\mbox{\scriptsize\boldmath${\cal N}$}}^2 + \fV - \fE\}\dot{\mbox{ }} = 
2(\mbox{\boldmath${\cal P}$}, \dot{\mbox{\boldmath${\cal P}$}})_{\mbox{\scriptsize\boldmath${\cal N}$}} + 
||\mbox{\boldmath${\cal P}$}||_{\dot{\mbox{\scriptsize\boldmath${\cal N}$}}}^2 
+ \dot{\fV} = ||\mbox{\boldmath${\cal P}$}||_{\mbox{\scriptsize\boldmath${\cal N}$}
                                              \dot{\mbox{\scriptsize\boldmath${\cal M}$}}
                                              \mbox{\scriptsize\boldmath${\cal N}$}}^2 - \dot{\fV} + 
||\mbox{\boldmath${\cal P}$}||_{\dot{\mbox{\scriptsize\boldmath${\cal N}$}}
                                \mbox{\scriptsize\boldmath${\cal M}$}
                                \mbox{\scriptsize\boldmath${\cal N}$}}^2 + \dot{\fV} = 
||\mbox{\boldmath${\cal P}$}||_{\{ \mbox{\scriptsize\boldmath${\cal NM}$} \}\dot{\mbox{ }}
                                   \mbox{\scriptsize\boldmath${\cal N}$}}^2 = 0
\eeq
by the chain rule and $\mbox{\boldmath${\cal NM}$} = \mathbb{I}$.

In the Appendix I compute ${\cal N}^{\bar{p}\bar{q}}$, ${\cal M}_{{\bar{p}\bar{q}},{\bar r}}$ and 
${\Gamma^{\bar{p}}}_{\bar{q}\bar{r}}$ 
for each mechanical theory, thus rendering the above equations explicitly-computed.

%========================================================================================================
\subsection{Mechanics on relative space and relational space}
%========================================================================================================

$E^2/2J = \dot{J}^2/8J$ is now not subtracted off in the action.  
Thus, on relative space,
\beq
\fT = \frac{1}{2}\left\{\frac{\dot{J}^2}{4J} + 
J\frac{\{1 + ||\mbox{\boldmath${\cal X}$}||^2 \} ||\dot{\mbox{\boldmath${\cal X}$}}||^2 - 
(\mbox{\boldmath${\cal X}$}\cdot\dot{\mbox{\boldmath${\cal X}$}})^2 }{\{1 + ||\mbox{\boldmath${\cal X}$}||^2 \}^2}
\right\}
\label{Lux}
\eeq
On 1d relational space is the same as relative space so one has a subcase of the above.  
On 2d relational space, one has 
\beq
\fT = \frac{1}{2}
\left\{
\frac{\dot{J}^2}{4J} + 
J\frac{    \{1 + ||\mbox{\boldmath${\cal Z}$}||_{\sC}^2\} 
                ||\dot{\mbox{\boldmath${\cal Z}$}}||_{\sC}^2 - 
                |(\mbox{\boldmath${\cal Z}$} \cdot \dot{\mbox{\boldmath${\cal Z}$}})_{\sC}|^2    }
     {    \{1 + ||\mbox{\boldmath${\cal Z}$}||_{\sC}^2\}^2    }
\right\}
\label{em}
\eeq
One can think of (\ref{Lux}) as a shape-scale split of the polar coordinates presentation
\beq 
2\fT = ||\dot{\bR}||_{\mbox{\scriptsize\boldmath${\mu}$}}^2 = 
||\mbox{\boldmath${\rho}$}||_{\mbox{\scriptsize\boldmath${\mu}$}}^2 + 
||\mbox{\boldmath$\rho\dot{\theta}$}||_{\mbox{\scriptsize\boldmath${\mu}$}}^2 \mbox{ } .  
\eeq
This is the Jacobi coordinates diagonal rearrangement of 
$\fT = \stackrel{\sum\sum}{\mbox{\scriptsize A $<$ B}}\frac{m_Am_B}{M}|\dot{r}_{AB}|^2$.  
One can think of (\ref{em}) similarly as the shape-scale split of the presentation 
\beq
2\fT = ||\mbox{\boldmath$\dot{\rho}$}||^2_{\mbox{\scriptsize\boldmath${\mu}$}} + 
||\mbox{\boldmath${\cal J}$}||^2_{\mbox{\scriptsize\boldmath${\cal I}$}} \mbox{ } , 
\eeq
which comes about by rearranging $\fT = \stackrel{\sum\sum}{\mbox{\scriptsize A $<$ B}}\frac{J_AJ_B}{J}|\theta_{AB}|^2$ 
for $J_A$ the Ath partial moment of inertia by introducing diagonalizing velocities ${\cal J}_{a}$ in 
analogy with the abovenoted Jacobi coordinates manipulation.  
The relational space presentations above are fully reduced reformulations of Barbour--Bertotti theory 
in 1d and 2d.

%========================================================================================================
%========================================================================================================
\section{Equivalence of my direct construction and the Barbour-type indirect construction}  
%========================================================================================================
%========================================================================================================

I now show that two of the above theories are reformulations of two of Barbour's mechanics theories.
Thus Kendall's work is (a good start on) the configuration space study for Barbour's mechanics;  
additionally, we have seen that the configuration spaces are $\mathbb{S}^k$ and $\mathbb{C}^k$ 
for these theories (in 1 or 2 dimensions).  As these are well-known spaces, this observation permits me 
to `complete' the configuration space study with the material provided in the Appendices.  

\mbox{ }  

\noindent{\bf Theorem 7} The following theories that I have built in this paper by the direct 
implementation of spatial relationalism are equivalent to the theories obtained by the Barbour-type 
indirect implementation of spatial relationalism, being related to these by the process of (Routhian) 
reduction. 

\noindent i) Theory 1 is equivalent to my new Theory (Example 5). 

\noindent ii) In 1d, Theory 2 is equivalent to Barbour's dilation-invariant Theory (Example 4)  

\noindent iii) In 2d, Theory 2 is equivalent to Barbour's dilation-invariant Theory.  

\noindent\underline{Proof}
i) For my new Theory as formulated by (\ref{9}) in (\ref{action}), 
the Lagrangian forms of the constraints (\ref{ZM}, \ref{ZDM}) are, by using (\ref{mom4}) and multiplying 
by $J\dot{\mbox{I}}$,
\beq
\underline{\ttP} = \sum_{{\cal A} = 1}^{{\cal N}} 
m_{\cal A}\delta^{\cal AB}\{\dot{\q}_{\cal B} - \dot{\a} + \dot{\C}\q_{\cal B}\} = 0 
\mbox{ } 
\eeq
and
\beq
{\ttD} = \sum_{{\cal A} = 1}^{{\cal N}} \q_{\cal A} \cdot 
m_{\cal A}\delta^{\cal AB}\{\dot{\q}_{\cal B} - \dot{\a} + \dot{\C}\q_{\cal B}\} = 0 
\mbox{ } .
\eeq
Then, using these equations and Routhian reduction to eliminate $\dot{\a}$ and $\dot{\C}$ from the 
action, I obtain the new action (\ref{action}) with  
\beq
\fT = \{PT - E^2\}/2J
\eeq
for
\beq 
J = ||\bR||^2_{\mbox{\scriptsize\boldmath${\mu}$}} \mbox{ } , \mbox{ }
T = ||\dot{\bR}||^2_{\mbox{\scriptsize\boldmath${\mu}$}} \mbox{ } , \mbox{ } 
P = ||\mbox{\boldmath{$\rho$}}||^2_{\mbox{\scriptsize\boldmath${\mu}$}} \mbox{ } \mbox{ and }
E = (\bR,\dot{\bR})_{\mu} \mbox{ } , 
\eeq 
for ${\mbox{\boldmath${\mu}$}}$ the array with components $\mu_{A}\delta_{AB}$ for $\mu_A$ the Jacobi 
cluster masses \cite{Marchal}.  
Dividing top and bottom by $R_1^4$ and making the identification 
\beq
\sqrt{\mu}_AR_{A\alpha} = {\cal X}_{A\alpha}
\eeq 
to pass to the Beltrami coordinates of Lemma 2, this is of the form (\ref{9}).  

\mbox{ }

\noindent ii) As in 1d the above theory coincides with Barbour's dilation-invariant Theory 
(through the absense of a continuous group of rotations in 1d) 
reduction to Barbour's dilation-invariant Theory in 1d is also accomplished as above.  

\mbox{ }  

\noindent iii) For Barbour's dilation-invariant Theory as formulated by (\ref{18b}, \ref{19b}) in 
(\ref{action}), the Lagrangian forms of the constraints (\ref{ZM}, \ref{ZAM} \ref{ZDM}) are, by using 
(\ref{mom4}) and multiplying by $J\dot{\mbox{I}}$,
\beq
\underline{\ttP} = \sum_{{\cal A} = 1}^{\N}{m_{\cal A}}\delta^{{\cal AB}}\{\dot{\q}_{\cal B} - 
\dot{\a} - \dot{\b} \cr \q_{\cal B}  + \dot{\C}\q_{\cal B}\}  = 0 \mbox{ } , \mbox{ } 
\eeq
\beq
\underline{\ttL} = \sum_{{\cal A} = 1}^{\N} \q^{\cal A} \cr {m_{\cal A}}\delta^{{\cal AB}}\{\dot{\q}_{\cal B} - \dot{\a} - 
\dot{\b} \cr \q_{\cal B}  + \dot{\C}\q_{\cal B}\} 
 = 0 \mbox{ } , \mbox{ } 
\eeq
and
\beq
\ttD \equiv \sum_{{\cal A} = 1}^{\N} \q^{\cal A} \cdot {m_{\cal A}}\delta^{{\cal AB}}\{\dot{\q}_{\cal B} - \dot{\a} - 
\dot{\b} \cr \q_{\cal B}  + \dot{\C}\q_{\cal B}\}  = 0 \mbox{ } .
\eeq

If the dimension is now furthermore 2, these equations and Routhian reduction can be used to 
eliminate $\dot{\a}$ and $\dot{\C}$ from the action.  
I thus obtain the new action (\ref{action})  
\beq
2J\fT = \{JT - E^2 - A\}/J  
\eeq
where  
\beq
A = \sum_I\sum_J\{(\R_I,\R_J)(\dot{\R}_I,\dot{\R}_J) - 
(\R_I,\dot{\R}_J)(\dot{\R}_I,\R_J)\} \mbox{ } . 
\eeq
I then recast this in terms of polar (rather than Cartesian) Jacobi coordinates, and then 
re-express these as complex variables, whereupon 
\beq
A = \sum_I\sum_J\rho_I^2\rho_J^2\dot{\theta}_I\dot{\theta}_J 
\eeq
and 
\beq 
E^2 = \sum_I\sum_J\rho_I\dot{\rho}_I\rho_J\dot{\rho}_J
\mbox{ } ,   
\eeq
so   
\beq
E^2 + A = |(\bar{\bz}\cdot\dot{\bz})_{\sC}|^2 \mbox{ } .  
\eeq 
Also, 
\beq 
J = |\bz|_{\sC}^2 \mbox{ } \mbox{ and } \mbox{ } T = |\dot{\bz}|_{\sC}^2 \mbox{ } . 
\eeq
So 
\beq 
JT - E^2 - A = |\bz|_{\sC}^2|\dot{\bz}|_{\sC}^2 - |(\bar{\bz} \cdot \dot{\bz})_{\sC}|^2 \mbox{ } . 
\eeq
Then 
\beq
{\fT} = 
\frac{1}{2}\frac{|\bz|_{\sC}^2|\dot{\bz}|_{\sC}^2 - |(\bar{\bz}\cdot\dot{\bz})_{\sC}|^2}
{\{|\bz|_{\sC}^2\}^2} \mbox{ } ,
\eeq  
which is the standard pre-Fubini form [c.f. first equality in (\ref{Easter})], which becomes the Fubini 
form (\ref{10}) upon division by the fourth power of one of the $z_i$ and adoption of the inhomogeneous 
coordinates ${\cal Z}_{\mbox{\scriptsize \tt a}}$ of Lemma 3 $\Box$.

%========================================================================================================
%========================================================================================================
\section{Conclusion}
%========================================================================================================
%========================================================================================================

This paper considers the topological and geometrical structure of the spaces of shapes with and 
without scale.  
Using this as first principles, I was led to various relational particle dynamics which have these 
spaces as their configuration spaces.  
I then found these to be equivalent to Barbour's relational particle dynamics \cite{B03} or other theories that 
are obtainable from Barbour's perspective once rather elaborate reduction has taken place.  
Thus Barbour's foundations are not the only way of thinking from which such theories can be extracted.

%%%%%%%%%%%%%%%%%%%%%%%ANSWER TO INTRODUCTION'S ISSUE 3
%
Another value of this paper is that one of the theories considered is new.  
It is a theory in which position and size are relational but orientation is absolute 
(while in Barbour's theory all three are relational).  
One use for this is as a simple model of whether Barbour's theory's dilation invariance models nature, 
Barbour's theory itself being hard to compute with in dimension 3 while my theory is equally simple 
in all dimensions.  
Thus my theory would allow for investigation of whether realistic -- i.e. 3d -- galactic and 
cosmological matter distributions can both reproduce known solar system physics and do better than 
Newtonian theory as regards modelling at galactic and cosmological scales.     
[The techniques of this paper would also permit to investigate Barbour theory proper -- but in 2-d -- 
in this regard, which itself may be reasonable for some rough calculations, as both the solar system 
and our galaxy are approximately planar.]

The principal purpose of this paper is that it provides a fairly full configuration space study for 
these theories.  
This mostly came about from my observation that these theories' configuration spaces coincide with 
the spaces studied disjointly in the geometrical and statistical literature, mostly by Kendall 
\cite{Kendall84, Kendallbook}.  

My theory in arbitrary dimension turns out to have configuration space $\mathbb{S}^{\mbox{\scriptsize N}\d - 1}$ for N 
$= \N - 1$, with standard spherical metric.    
Barbour's theory coincides with mine in 1d so that has configuration space $\mathbb{S}^{\mbox{\scriptsize N} - 1}$ too.  
Barbour's theory in 2-d turns out to have the configuration space $\mathbb{CP}^{\mbox{\scriptsize N} - 1}$ with standard 
Fubini--Study metric.    
These are all well-known topologies and geometries, so making these identifications provides a wealth 
of techniques and results for the study of relational models at both classical and quantum levels, which 
are of interest in the investigations detailed in the following paragraphs.
This paper is in this respect a substantial improvement over the study of configuration spaces 
`from scratch' in \cite{06I} and which does not explicitly complete the reduction process, and over 
the study in \cite{07I} (which is for only 3 particles); the `simple ratio variables' of 
\cite{06II, 07I} are now recognized as the inhomogeneous coordinates of projective geometry. 
The extension to more than 3 particles in  2-d is a significant generalization in terms of 
containing two non-trivial subsystems and in having non-conformally flat configuration space geometry.  
I have also touched on formulations of the mechanical theory with relative position and orientation and 
absolute scale (Barbour--Bertotti theory \cite{BB82}) in shape-and-scale variables in 1 and 2 dimensions.  
I speculate that Kendall's work on higher-d shape spaces could play an analogous role in higher-d Barbour theory and Barbour--Bertotti theory to that 
which his 1 and 2 dimensional work plays in this paper.  
This further work of Kendall's includes a considerable study at the topological level \cite{Kendallbook}, 
and quite a complete study of further structures in the case of 4 particles in 3-d \cite{Kendallbook, Kendall43}.   
Both the geometry and the mechanics being harder in the 3-d case, I leave this study for a future occasion.

%%%%%%%%%%%%%%%%%%%%% ANSWERS 1) OF INTRO
%
In this paper's study of Barbour's theory in 2-d, I observe the continuation of the trend \cite{06I, 07I} 
that the relational approach gives simple mathematics similar to that encountered in absolute approach.  
 For d = 3, the simpleness disappears.  
Barbour's theory may be relatively simpler to handle than Barbour--Bertotti theory because it avoids 
having a maximal collision.  
This may make Barbour theory relevant as the simplest mathematics that arises relationally and is 
substantially different from the mathematics of absolute mechanics.
\cite{07I} is about triangle lands, of interest as in \cite{B94I, EOT}.  
This paper is about polygon lands in 2d, of interest likewise.  
One direction for further research is to solve the classical equations of motion for $\N > 3$ in 
parallel to the $\N = 3$ treatment in \cite{07I}.

%%%%%%%%%%%%%%%%%%%%%%% ANSWERS 2) OF INTRO
%
As regards quantum relational mechanics, in the light of this paper one can now use known approaches to 
quantization on the configuration spaces $\mathbb{S}^k$ and $\mathbb{CP}^k$, but now interpreting these 
in the new, relational context by mapping back from shape coordinates to particle position coordinates. 
I set this up in \cite{07II}, additionally providing exact, perturbative and numerical solutions for 
various natural potentials in the simplest 2-d case (by using $\mathbb{CP}^1 = \mathbb{S}^2$). 
Lines of work that remain to be done are explicitly solving some cases of QM for the 2-d 4-particle 
$\mathbb{CP}^2$ configuration space system, and considering cases 
whose potentials are such that relative angular momentum exchange between subsystems \cite{06I, 07I} 
is possible.

%%%%%%%%%%%%%%%%%%%%%%% ADDRESSES THE MAIN ISSUE IN THE INTRODUCTION
%
As argued in the Introduction, relational particle dynamics are toy models that are useful as regards 
the investigation of conceptual issues in the problem of time in quantum GR. 
The cases studied in this paper are {\sl not} good as toy models for the study of superspace (a configuration space for GR), 
because their configuration spaces are too simple to manifest most of the difficulties that one encounters 
in studying superspace.  
Rather, the relational models studied in this paper are toy models for what one could do as regards 
the problem of time in GR were its configuration space substantially mastered.

One problem of time approach is {\it records theory}.  
This is at present heterogenous rather than a single subject, with Page--Wootters \cite{PW83},  
Gell-Mann--Hartle \cite{GMH} and Halliwell \cite{H99}, and Barbour \cite{B94II, EOT} covering different 
suggestive aspects.  
I take it to mean the study of correlations between subsystems of a single present in the hope 
of being able to reconstruct (something) of a semblance of dynamics or history from them \cite{ARec}.
A useful tool for this would be a minimizer to compare different (sub)configurations \cite{ARec}, 
in which respect I observe that Kendall's minimizer (\ref{KenMin}) is a candidate in addition to 
Barbour's [minimize the redundant form of the relational particle model's action over the auxiliary 
variables].  
See \cite{Dist} for more.    
These tools are directly applicable at the classical level (for which records theory already makes sense) 
and may be inherited at (or induce more complicated structures at) the quantum level (for which 
records theory is likely to be more interesting).  
Barbour conjectures \cite{B94II, EOT} that records which give a sembance of dynamics, which he terms 
{\it time capsules} and include bubble chambers which exhibit the tracks of $\alpha$-particles 
\cite{Mott}, are states in geometrically special parts of the configuration space for which the 
wavefunction of the universe is highly peaked.\foo{That 
%%%%%%%%%%%%%%%%%%%%%%%%%%%%%%%%%%%%%%%%%%%%%%%%%%%%%%%%%%%%%%%%%%%%%%%%%%%%%%%%%%%%%%%%%%%%%%%%%%%%%%%%%
this is relevant to quantum cosmology: e.g. 
Halliwell and Castagnino--Laura \cite{Hallioverlap, HaLaCa} have considered quantum cosmology in broadly 
similar terms to this $\alpha-particle$ paradigm, as well as records theory 
and its semblance of time being, at least conceptually, a way of resolving the problem of time.}
%%%%%%%%%%%%%%%%%%%%%%%%%%%%%%%%%%%%%%%%%%%%%%%%%%%%%%%%%%%%%%%%%%%%%%%%%%%%%%%%%%%%%%%%%%%%%%%%%%%%%%%%%
%
%In this respect, I observe that Kendall also supplies statistical tools on shape spaces for deciding 
%whether such as approximate collinearities are accidental or significant \cite{Kendall84, Kendallbook}.***
%
Conformal non-flatness is needed for the fully unambiguously geometrical kinetic term effects suggested by Barbour 
in \cite{EOT}.
This paper makes clear that the simplest such relational dynamics is 4 particles in 2d (configuration 
space $\mathbb{CP}^2$) , which lies within 
the scope of the relational program via this paper's results.    
This model is also motivated by nontrivial records theory requiring models that have enough degrees 
of freedom to permit two spatially-separated nontrivial subsystems.

This paper also allows for more advanced specific toy modelling of the emergent semiclassical time 
approach than that of the specific semiclassical examples in \cite{SemiclI}. 
The $\mathbb{S}^2 = \mathbb{CP}^1$ case provides an underlying geometrical understanding for the case 
with nontrivial linear constraints and 2 heavy slow variables and 2 light fast variables in the unreduced 
picture.   
This is used in \cite{SemiclIII} as a toy model exhibiting some of the 
complications of inhomogeneous perturbations about minisuperspace and of some gravity--matter systems.    
The theory on $\mathbb{CP}^2$ allows for the study of 2 heavy slow variables and two nontrivial subsystems of 2 light 
fast degrees of freedom each and hence is a simple model of studying correlations between two 
bumps in an expanding universe cosmology -- i.e. it is a toy model of the situation studied by 
Halliwell and Hawking \cite{HallHaw}.

The counterpart of whether microsuperspace dynamics lies stably within minisuperspace dynamics 
\cite{KR89} is also a potential application of this paper.  
E.g. whether the dynamics of a $\mathbb{CP}^1$ subsystem lies stably within a $\mathbb{CP}^2$ subsystem, 
with the benefit of here being able to extend the study to a readily available and natural nested 
hierarchy of $\mathbb{CP}^k$ subsystems.     
This paper would also be useful as regards studying internal time [provided by $(\bR, \bP)$ in 
theories with scale] in more complicated examples than those  in \cite{06II}.  
%
%Could also do histories theory in this setting but do not currently have any motivation for this.  

\mbox{ }

%========================================================================================================
\noindent{\bf Acknowledgments}
%========================================================================================================

\mbox{ }

\noindent I thank Dr. Julian Barbour for the idea of working on relational particle models with and 
without scale, discussions, hospitality and for reading the manuscript.  
I thank the Lecturers responsible for the advanced geometrical courses in the Cambridge Mathematics 
Tripos over the past 10 years, which helped me with techniques and understanding.  
I thank Professor Don Page for previous discussions and references about spheres and complex 
projective spaces which once again proved useful.  
I thank Dr Fay Dowker for having me give a lunchtime seminar/discussion at Imperial 
College London on an earlier draft of this article.  
And I thank Peterhouse Cambridge for funding me through a Research Fellowship in 2006--2008.    

\mbox{ }

%========================================================================================================
\noindent{\bf Appendix A: geometrical properties of RPM configuration spaces}
%========================================================================================================

\mbox{ }

\noindent In order for this Paper's equations to be explicit, I need the following objects.  

For $\fP(\N, \d) = \mathbb{S}^{\sn\sd - 1}$, the inverse metric is 
\beq
{\cal N}^{\Gamma\Delta} =  \{1 + ||{\cal X}||^2\}  \{
\delta^{\Gamma\Delta} + {\cal X}^{\Gamma} {\cal X}^{\Delta}\} \mbox{ } .
\eeq
The first partial derivatives of the metric are 
\beq
{\cal M}_{\Gamma\Delta, \Lambda} = - \frac{        
{\cal X}_{\Gamma} \delta_{\Delta\Lambda} + {\cal X}_{\Delta} \delta_{\Gamma\Lambda}   + 
2{\cal X}_{\Lambda} \delta_{\Gamma\Delta}    }{    \{1 + ||{\cal X}||^2\}^2    }      + 
4\frac{{\cal X}_{\Gamma}{\cal X}_{\Delta}{\cal X}_{\Lambda}}{\{1 + ||{\cal X}||^2\}^3} \mbox{ } .  
\eeq
The Christoffel symbols are 
\beq
{\Gamma^{\Lambda}}_{\Gamma\Delta} = - \frac{        
{\cal X}_{\Gamma} {\delta_{\Delta}}^{\Lambda} + {\cal X}_{\Delta} {\delta_{\Gamma}}^{\Lambda}     }
{    1 + ||{\cal X}||^2   } \mbox{ } .
\eeq
These spaces have Ricci tensor
\beq
R_{\Gamma\Delta} = \{\n\d - 2\}{\cal M}_{\Gamma\Delta}
\eeq
(so $\mathbb{S}^{\sn\sd - 1}$ is Einstein) and hence have constant Ricci scalar curvature
\beq
R = \{\n\d - 1\}\{\n\d - 2\} \mbox{ } .  
\eeq 
These spaces are all conformally flat, as an easy consequence of their being maximally symmetric. 
[They have Nd(Nd - 1)/2 Killing vectors.]

For $\fS(\N, \d) = \mathbb{CP}^{\sn - 1}$, write the metric in 2 blocks, 
\beq
{\cal M}_{pq} = \{1 + ||{\cal R}||^2\}^{-1}\delta_{pq} - \{1 + ||{\cal R}||^2\}^{-2}{\cal R}_p{\cal R}_q 
\mbox{ } ,
\eeq 
\beq
{\cal M}_{\tip\tiq} = \{\{1 + ||{\cal R}||^2\}^{-1}\delta_{\tip\tiq} - 
\{1 + ||{\cal R}||^2\}{\cal R}_{\tip}{\cal R}_{\tiq}\} {\cal R}_{\tip}{\cal R}_{\tiq}
\mbox{   (no sum) ,}
\eeq 
\beq
{\cal M}_{p\tiq} = 0 \mbox{ } .  
\eeq
Then the inverse metric is 
\beq
{\cal N}^{pq} = \{1 + ||{\cal R}||^2\}\{\delta^{pq} + {\cal R}^p{\cal R}^q\} 
\eeq
\beq
{\cal N}^{\tip\tiq} = \{1 + ||{\cal R}||^2\}\{\delta^{\tip\tiq}/{\cal R}^2_{\tip} + 1^{\tip\tiq}\}  
\mbox{ (no sum) ,}
\eeq
for $1^{\tip\tiq}$ the matrix whose entries are all 1, and   
\beq
{\cal N}^{p\tiq} = 0 \mbox{ } .  
\eeq
Then the only nonzero first partial derivatives of the metric are (no sum)
\beq
{\cal M}_{pq,r} = \frac{1}{\{1 + ||{\cal R}||^2\}^2}
\left\{
\frac{4{\cal R}_{p}{\cal R}_{q}{\cal R}_{r}}{1 + ||{\cal R}||^2} - 
\{2{\cal R}_r\delta_{pq} + {\cal R}_q\delta_{pr} + {\cal R}_p\delta_{qr} \}
\right\} \mbox{ } , 
\eeq
\beq
{\cal M}_{\tip\tiq,r} = \frac{2{\cal R}_{\tip}{\cal R}_{\tiq}}{\{1 + ||{\cal R}||^2\}^2}
\left\{
\frac{2{\cal R}_{\tip}{\cal R}_{\tiq}{\cal R}_{r}}{1 + ||{\cal R}||^2} - 
\{{\cal R}_r\delta_{\tip\tiq} + {\cal R}_{\tiq}\delta_{\tip r} + {\cal R}_{\tip}\delta_{\tiq r} \}
\right\} + \frac{\delta_{\tip\tiq}\{{\cal R}_{\tip}\delta_{\tiq r} + {\cal R}_{\tiq}{\delta}_{\tip r}
\}}{1 + ||{\cal R}||^2} \mbox{ } .  
\eeq
The only nonzero Christoffel symbols are (no sum except over $\tilde{s}$) 
\beq
{\Gamma^{p}}_{qr} = - \frac{  {\cal R}_r{\delta^p}_q + {\cal R}_q{\delta^p}_r  }{1 + ||{\cal R}||^2} 
\mbox{ } ,
\eeq
\beq
{\Gamma^{p}}_{\tiq\tir} = \delta_{\tiq\tir}{\delta_{\tir}}^p{\cal R}_{\tiq} - 
{\cal R}_{\tiq}{\cal R}_{\tir}
\frac{  {\cal R}_{\tir}{\delta^p}_{\tiq} + {\cal R}_{\tiq}{\delta^p}_{\tir}  }{1 + ||{\cal R}||^2} 
\eeq 
and
\beq
{\Gamma^{\tip}}_{\tiq r} = \{\delta^{\tip\tis}/{\cal R}^2_{\tip} + 1^{\tip\tis}\}
\left\{\frac{{\cal R}_{\tis}{\cal R}_{\tiq}}{\{1 + ||{\cal R}||^2\}^2}
\left\{
\frac{2{\cal R}_{\tis}{\cal R}_{\tiq}{\cal R}_{r}}{1 + ||{\cal R}||^2} - 
\{{\cal R}_r\delta_{\tiq\tis} + {\cal R}_{\tis}\delta_{\tiq r} + {\cal R}_{\tiq}\delta_{\tis r} \}
\right\} + \frac{\delta_{\tis\tiq}\{{\cal R}_{\tis}\delta_{\tiq r} + {\cal R}_{\tiq}{\delta}_{\tis r}
\}}{1 + ||{\cal R}||^2}\right\} \mbox{ } .  
\eeq
These spaces have Ricci tensor 
\beq
R_{\sa\sb} = 2\n {\cal M}_{\sa\sb}
\eeq 
(so $\mathbb{CP}^{\sn - 1}$ is Einstein) and thus these are also spaces of constant Ricci scalar curvature 
\beq
R = 4\n\{\n - 1\} \mbox{ } .  
\eeq 
However, for $\N > 3$, they have nonzero Weyl tensor (as checked by Maple \cite{Maple}) 
and so are not conformally flat.  [They are fairly symmetrical but not maximally symmetric for 
$\N > 1$, e.g. $\mathbb{CP}^2$ has 8 Killing vectors \cite{GiPo}.]

None of the abovementioned curvatures, or curvature scalars constructed from them and the metric, 
blow up for finite ${\cal R}_a$

The $\mathbb{S}^{\sk}$ are real manifolds.  
The $\mathbb{CP}^{\sk}$ are complex manifolds.  
$\mathbb{S}^{2} = \mathbb{CP}^1$ is the only complex manifold among the $\mathbb{S} ^{\sk}$.  
The $\mathbb{CP}^{\sk}$ are, furthermore, K\"{a}hler, with K\"{a}hler potential  
$K = \sum_{A = 1}^{\sm}|{\cal Z}_A|^2$.  
The Euler and Pontrjagin classes of those which are real manifolds and the 
Chern classes and characters of those which are complex manifolds are readily computible (these are 
defined in e.g. \cite{Nakahara} and are important as obstructions to quantization, and as regards 
issues concerning instantons and magnetic charges).    

\mbox{ }

%========================================================================================================
\noindent{\bf Appendix B: topological properties of RPM configuration spaces}
%========================================================================================================

\mbox{ }

\noindent Now that $\fP({\cal N}, \md)$, $\fS({\cal N}, 1)$ and $\fS({\cal N}, 2)$ have been identified 
as $\mathbb{S}^{\sk}$ and $\mathbb{CP}^{\sk}$ manifolds, the following classically and 
quantum-mechanically useful topological information about paths on and obstructions in these 
configuration spaces becomes available.

$\fP({\cal N}, \d) = \mathbb{S}^{Nd - 1}$, $\fS({\cal N}, 1) = \fP({\cal N}, 1)$ and 
$\fS({\cal N}, 2) = \mathbb{CP}^{{\cal N} - 2}$ are compact without boundary and Hausdorff.

Following from \cite{Hatcher}, the homotopy groups of $\fP({\cal N}, \d)$ exhibit the simple patterns  

\noindent $\pi_{\sp}(\fP(3, 1)) = \pi_{\sp}(\mathbb{S}^{1}) = 
\left\{
\stackrel{    \mbox{$\mathbb{Z}$}    }{    0    }
\stackrel{    \mbox{p $= 1$}    }{    \mbox{ otherwise    }}
\right. , 
\pi_{\sp}(\fP({\cal N}, \d)) = \pi_{\sp}(\mathbb{S}^{\sN\d - 1}) = 
\left\{
\stackrel{    \mbox{$\mathbb{Z}$}    }{    0    }
\stackrel{    \mbox{ } \mbox{ p = Nd -- 1 $> 1$}    }{  \mbox{ } \mbox{ p $ < $ Nd -- 1} } 
\right.$ 

\noindent From \cite{Toda}, the first few homotopy groups in the remaining wedge are

\noindent
\begin{tabbing}
                 \hspace{1.2in}                 \= 
                 \hspace{0.2in}                 \=
$\pi_3$          \hspace{0.4in}                 \=
$\pi_4$          \hspace{0.4in}                 \=
$\pi_5$          \hspace{0.4in}                 \=
$\pi_6$          \hspace{0.4in}                 \=   
$\pi_7$          \hspace{0.4in}                 \=
$\pi_8$          \hspace{0.4in}                 \=
$\pi_9$          \hspace{0.4in}                 \=   
$\pi_{10}$       \hspace{0.4in}                 \=
$\pi_{11}$       \hspace{0.4in}                 \=     \\ 
\fP(4, 1) = \fS(3, 2) = \> $\mathbb{S}^{2}$     \>
$\mathbb{Z}$                                    \>
$\mathbb{Z}_2$                                  \>
$\mathbb{Z}_2$                                  \>
$\mathbb{Z}_{12}$                               \>
$\mathbb{Z}_{2}$                                \>
$\mathbb{Z}_{2}$                                \>
$\mathbb{Z}_{3}$                                \>
$\mathbb{Z}_{15}$                               \>
$\mathbb{Z}_{2}$                                \>   \\
\fP(5, 1) = \fP(3, 2) = \> $\mathbb{S}^{3}$     \>
                                                \>
$\mathbb{Z}_2$                                  \>
$\mathbb{Z}_2$                                  \>
$\mathbb{Z}_{12}$                               \>   
$\mathbb{Z}_{2}$                                \>
$\mathbb{Z}_{2}$                                \>
$\mathbb{Z}_{3}$                                \>
$\mathbb{Z}_{15}$                               \>
$\mathbb{Z}_{2}$                                \>   \\
\fP(6, 1) = \> $\mathbb{S}^{4}$                 \>
                                                \>
                                                \>
$\mathbb{Z}_2$                                  \>
$\mathbb{Z}_2$                                  \>
$\mathbb{Z} \times \mathbb{Z}_{12}$             \>
$\mathbb{Z}_{2} \times \mathbb{Z}_2$            \> 
$\mathbb{Z}_{2} \times \mathbb{Z}_2$            \>
$\mathbb{Z}_{24} \times \mathbb{Z}_3$           \>
$\mathbb{Z}_{15}$                               \>    \\
\fP(7, 1) = \fP(4, 2) = \> $\mathbb{S}^{5}$     \>
                                                \>
                                                \>
                                                \>
$\mathbb{Z}_2$                                  \>
$\mathbb{Z}_{2}$                                \>
$\mathbb{Z}_{24}$                               \>
$\mathbb{Z}_{2}$                                \>
$\mathbb{Z}_{2}$                                \>
$\mathbb{Z}_{2}$                                \>    \\
\fP(8, 1) = \> $\mathbb{S}^{6}$                 \>
                                                \>
                                                \>
                                                \>
                                                \>
$\mathbb{Z}_2$                                  \>
$\mathbb{Z}_{2}$                                \>
$\mathbb{Z}_{24}$                               \>
0                                               \>
$\mathbb{Z}$                                    \>    \\
\fP(9, 1) = \fP(5, 2) = \> $\mathbb{S}^{7}$     \>
                                                \>
                                                \>
                                                \>
                                                \>
                                                \>
$\mathbb{Z}_2$                                  \>
$\mathbb{Z}_{2}$                                \>
$\mathbb{Z}_{24}$                               \>
0                                               \>    \\
\end{tabbing}

From \cite{Hatcher}, it follows that the homology and cohomology groups are  

\noindent $H_{\sp}(\fP({\cal N}, \d)) = H_{\sp}(\mathbb{S}^{\sN\sd - 1}) = 
\left\{ 
\stackrel{    \mbox{$\mathbb{Z}$}    }{    0    }  
\stackrel{    \mbox{ } \mbox{ p $= 0$ or Nd - 1}    }{   \mbox{ } \mbox{ otherwise}    } 
\right\} 
= H^{\sp}(\fP({\cal N}, \d)) = H^{\sp}(\mathbb{S}^{{\sN\sd}- 1})$.

The first and second Stiefel--Whitney classes are trivial for all spheres \cite{Nakahara}, which imply 
respectively that $\fP({\cal N}, \d)$ are all orientable and admit a nontrivial spin structure.

As $\fS({\cal N}, \d) = \fP({\cal N}, \d)$, one can read off the corresponding results for 
$\fS({\cal N}, \d)$ from the above by setting d = 1.

From \cite{Whitehead}, it follows that the homotopy groups 
$\pi_{\sp}(\fS({\cal N}, 2)) = \pi_{\sp}(\mathbb{CP}^{{\cal N} - 2}) = 
\left\{ \stackrel{\mbox{$\mathbb{Z}$}}{\pi_{\sp}(\mathbb{S}^{2{\cal N} - 3})}\right. 
\stackrel{\mbox{p = 2}}{\mbox{otherwise}}$.

From \cite{Hatcher}, it follows that the homology and cohomology groups are 

\noindent $H_{\sp}(\fS({\cal N}, 2)) = H_{\sp}(\mathbb{CP}^{{\cal N} - 2}) = 
\left\{ 
\stackrel{   \mbox{$\mathbb{Z}$}    }{    0    }  
\stackrel{   \mbox{ } \mbox{ p even up to $2\{{\cal N} - 2\}$}    }{  \mbox{ }  \mbox{ otherwise}    } 
\right\} 
= H^{\sp}(\fS({\cal N}, 2)) = H^{\sp}(\mathbb{CP}^{{\cal N} - 2})$.

The first Stiefel--Whitney classes are trivial for all complex projective spaces \cite{Nakahara}, 
which implies that all the $\fS({\cal N}, 2)$ are orientable.  
The second Stiefel--Whitney classes are trivial for ${\cal N} - 2$ an odd integer 
(so these $\fS({\cal N}, 2)$ admit all a nontrivial spin structure), and are nontrivial 
[equal to the generator of $H^2(\mathbb{CP}^{{\cal N} - 2}, \mathbb{Z})]$ for ${\cal N} - 2$ an even 
integer (so that nontrivial spin structures do not exist for these due to topological obstruction).

%=====================================================BIBLIOGRAPHY==========================================================================

\end{document}